\documentclass[12pt]{article}
\usepackage{epsfig,amssymb,euscript}
\usepackage{amsmath,amscd}

\numberwithin{equation}{section}
\newcommand{\be}{\begin{eqnarray}}
\newcommand{\ee}{\end{eqnarray}}
\newcommand{\bea}{\begin{eqnarray}}
\newcommand{\eea}{\end{eqnarray}}
\newcommand{\ba}{\begin{array}}
\newcommand{\ea}{\end{array}}

\newcommand{\nn}{\nonumber \\}

\DeclareMathOperator{\SO}{\mathit{SO}}
\DeclareMathOperator{\SU}{\mathit{SU}}
\DeclareMathOperator{\Symp}{\mathit{Sp}}

\newcommand{\rep}[1]{\mbox{\boldmath${#1}$}}

\newcommand{\id}{\mathbf{1}}

\DeclareMathOperator{\Tr}{Tr}
\DeclareMathOperator{\diag}{diag}

\DeclareMathOperator{\Ricci}{\textit{Ricci}}

\newcommand{\bZ}{\mathbb{Z}}
\newcommand{\bR}{\mathbb{R}}

\newcommand{\bCP}{\mathbb{C}P}

\def\a{\alpha}   \def\b{\beta}   \def\l{\lambda}  \def\g{\gamma}
    \def\d{\delta}  \def\t{\theta}   \def\s{\sigma}
\def\e{\epsilon} 

 \def\L{\Lambda} \def\G{\Gamma}

 \newcommand{\vp}{\varphi}

\def\CL{{\cal L}}

\def\CM{{\cal M}}

\newcommand{\bref}[1]{(\ref{#1})}

\newcommand{\ben}{\begin{displaymath}}
\newcommand{\een}{\end{displaymath}}
\newcommand{\bean}{\begin{eqnarray*}}
\newcommand{\eean}{\end{eqnarray*}}
\newcommand{\bi}{\begin{itemize}}
\newcommand{\ei}{\end{itemize}}

\newcommand{\ct}{T_\mt{deconf}}

\def\l{\lambda}
\def\a{\alpha}

\def\b{\beta}
\def\g{\gamma}
\def\G{\Gamma}
\def\d{\delta}
\def\s{\sigma}
\def\e{\epsilon}

\def\vp{\varphi}
\def\vph{\varphi}

\def\otaula{\begin{tabular}}
\def\ctaula{\end{tabular}}

\renewcommand{\L}{\Lambda}
\renewcommand{\t}{\theta}

\def\bnum{\begin{enumerate}}
\def\enum{\end{enumerate}}

\def\CR{\mathbb{R}}
\def\CM{\mathcal{M}}
\def\CC{\mathbb{C}}

\def\s{\sigma}
\def\8M{$\CM_8$}

\def\be{\begin{equation}}
\def\ee{\end{equation}}
\def\G{\Gamma}
\def\g{\gamma}
\def\ei{e^{\underline{i}}}

\def\e1{e^{\underline{1}}}
\def\1u{\underline{1}}
\def\2u{\underline{2}}

\def\0u{\underline{0}}
\def\e{\epsilon}
\def\target{$\CR^{1,1}\times \mathcal{M}_8$ }
\def\target2{$\CR^{1,1}\times \mathcal{M}_8$,}
\def\9G{\G_{\underline{9}}}

\def\a{\alpha}
\def\b{\beta}
\def\undos{\frac{1}{2}}

\def\we{\wedge}

\newcommand{\caln}{\mbox{${\cal N}$}}

\newcommand{\pa}{\partial}

\newcommand{\sac}{\, , \qquad}

\def\w{\omega}
\def\r{\rho}

\def\m{\mu}
\def\n{\nu}
\def\half{\frac{1}{2}}
\def\goto{\rightarrow}

\def\cm{c_{\mu}}
\def\cm{c_{\mu}}
\def\sm{s_{\mu}}
\def\cmh{c_{\mu / 2}}
\def\smh{s_{\mu /2}}
\def\ct{c_{\theta}}
\def\st{s_{\theta}}
\def\ctp{c_{\theta'}}
\def\stp{s_{\theta'}}
\def\cps{c_{4\psi_-}}
\def\sps{s_{4\psi_-}}

\begin{document}



\begin{titlepage}

\vfill

\begin{flushright}
Imperial/TP/050502\\ hep-th/0505207\\
\end{flushright}
\vfill

\begin{center}

\baselineskip=16pt

{\Large\bf Marginal Deformations of Field Theories\\*[5pt]
  with $AdS_4$ Duals}

\vskip 1.3cm

Jerome P. Gauntlett, Sangmin Lee, Toni Mateos and Daniel Waldram

\vskip 1cm
{\small{\it
Blackett Laboratory, Imperial College\\ Prince Consort Rd\\London, SW7 2AZ, U.K\\}}

\end{center}

\vfill

\begin{center}
\textbf{Abstract}
\end{center}

\begin{quote}
We generate new $AdS_4$ solutions of $D=11$ supergravity starting
from $AdS_4\times X_7$ solutions where $X_7$ has $U(1)^3$ isometry.
We consider examples where $X_7$ is weak $G_2$, Sasaki-Einstein or
tri-Sasakian, corresponding to $d=3$ SCFTs with $\caln=1,2$ or 3
supersymmetry, respectively, and where the deformed solutions preserve
$\caln=1,2$ or 1 supersymmetry, respectively. For the special cases
when $X_7$ is $M(3,2)$, $Q(1,1,1)$ or $N(1,1)_I$ we identify the
exactly marginal deformation in the dual field theory. We also show
that the volume of supersymmetric 5-cycles of $N(1,1)_I$ agrees with
the conformal dimension predicted by the baryons of the dual field
theory.

\end{quote}

\vfill

\end{titlepage}

\section{Introduction}
It is not uncommon for supersymmetric conformal field theories to have exactly
marginal deformations that preserve the superconformal symmetry.
If a conformal field theory with exactly marginal operators has an $AdS$ dual \cite{Maldacena:1997re}
one should be able to construct whole families of $AdS$ solutions. An early perturbative construction
of such solutions was undertaken in \cite{Aharony:2002hx} for certain deformations of $AdS_5\times S^5$ .
In a recent work Lunin and Maldacena discovered a remarkably simple method for
generating new $AdS$ solutions (not necessarily supersymmetric)
starting from an $AdS$ solution with an internal space possessing
certain isometries \cite{Lunin:2005jy}. The new solutions describe
a special class of exactly marginal deformations of the CFT dual
to the original $AdS$ solution.

One focus of \cite{Lunin:2005jy} was
type IIB $AdS$ solutions with the internal space having a
$U(1)^2$ isometry. Such solutions can also be
viewed as solutions of the $D=8$ supergravity theory obtained by
compactifying type IIB on the two-torus. The duality group of this
$D=8$ supergravity is $SL(3,\bR)\times SL(2,\bR)$ (of which an
$SL(3,\bZ)\times SL(2,\bZ)$ subgroup survives in string theory). The
key observation of \cite{Lunin:2005jy} was that starting from a
given $AdS$ solution there are certain $SL(2,\bR)\subset
SL(3,\bR)$ transformations that generate new {\it regular} $AdS$
solutions. Furthermore, the corresponding exactly marginal
deformations (``$\beta$-deformations'') in the dual CFT were
identified. The techniques were illustrated with several $AdS_5$
examples, including $AdS_5\times S^5$, $AdS_5\times T^{1,1}$ and
$AdS_5\times Y^{p,q}$ where
$Y^{p,q}$ are the new Sasaki-Einstein spaces discovered in
\cite{Gauntlett:2004zh,Gauntlett:2004yd}. The $Y^{p,q}$ spaces have
recently been generalised in \cite{Cvetic:2005ft}. These new metrics
have $U(1)^3$ isometry, of which a $U(1)^2$ preserves the Killing
spinors, and we note that it is straightforward find the corresponding
supersymmetric deformed solutions. Deformations of non-conformal
geometries were also considered and a generalisation recently appeared
in \cite{Gursoy:2005cn}.

Lunin and Maldacena also generalised the technique to $AdS$
solutions of $D=11$ supergravity when the internal space has a
$U(1)^3$ isometry. These solutions are also solutions of the $D=8$ supergravity
theory obtained by compactifying $D=11$ supergravity on a
three-torus, which is in fact the same
$D=8$ supergravity theory arising from the compactification of type
IIB on a two-torus. In this case
the key duality transformations lie in the explicit $SL(2,\CR)$
factor in  $SL(3,\CR)\times SL(2,\CR)$. The technique was
illustrated using the maximally supersymmetric $AdS_4\times S^7$
solution.

Here we will study some additional $AdS_4$ examples in
$D=11$. The solutions that we use to generate the new solutions
describing the deformations are all of the Freund-Rubin form:
$AdS_4\times X_7$. Such solutions arise as the near horizon limit of
M2-branes lying at the singular apex of an eight-dimensional
Ricci-flat cone whose base is $X_7$. We will focus on
supersymmetric solutions in which case the cone must have special
holonomy \cite{Klebanov:1998hh,Acharya:1998db,Morrison:1998cs}, or equivalently $X_7$ admits at least one Killing spinor.
In order to find deformed solutions using the techniques of
\cite{Lunin:2005jy}, we need examples where $X_7$ has $U(1)^3$
isometry. If we demand that the deformed geometry preserves some
supersymmetry, which we mostly will, this $T^3$ action must preserve
at least one Killing spinor on $X_7$.

We will find deformations for a number of different $X_7$, which we
will review in section 2. Specifically we will consider the
tri-Sasakian metric on $N(1,1)$, which we denote by $N(1,1)_I$, the
Sasaki-Einstein metrics on $Q(1,1,1)$ and $M(3,2)$ along with their
co-homogeneity one generalisations found in
\cite{Gauntlett:2004hh}. We will also discuss the weak $G_2$ metrics
on the squashed seven-sphere and on $N(k,l)$.

It is interesting that dual $d=3$ conformal field theories with
$\caln=2$ supersymmetry have been proposed for the $Q(1,1,1)$ and
$M(3,2)$ solutions in \cite{Fabbri:1999hw} and with $\caln=3$
supersymmetry for the tri-Sasaki metric $N(1,1)_I$ in
\cite{Billo:2000zr} (see also \cite{Gukov:1999ya}),
and this will also be reviewed in section 2.
Although these field theories are still not well understood, a
spectrum of chiral operators consistent with the Kaluza-Klein
spectrum can be obtained. For the $Q(1,1,1)$ and $M(3,2)$ cases
the volumes of some supersymmetric 5-cycles have been shown to
agree with the conformal dimensions of baryon operators. The latter
calculation has not yet been performed for the $N(1,1)_I$ case so,
as somewhat of an aside, we do this in an appendix finding
results consistent with the field theory proposed in
\cite{Billo:2000zr}. In addition, for all three of these examples
we are able to identify the exactly marginal operators in the
field theories corresponding to the deformed supergravity solutions
that we construct.

The plan for the remainder of the paper is as follows. In section 3 we discuss
the technique of constructing the deformed solutions and analyse how
much of the supersymmetry is preserved. In section 4 we present the new deformed
supergravity solutions. In section 5 we analyse the field theory duals.
We conclude in section 6, which includes a discussion of some simple non-supersymmetric generalisations.

\section{Review of some $AdS_4\times X_7$ solutions and their duals}

Let us begin by recalling some well known facts about $AdS_4\times X_7$ solutions of
$D=11$ supergravity with $X_7$ an Einstein manifold and $F_4\propto vol_{AdS_4}$, where
$vol_{AdS_4}$ is the volume-form of $AdS_4$.
We choose the metric on $X_7$ to be normalised so that $Ricci(X_7)=6g(X_7)$,
just as for a unit radius $S^7$.
In order to preserve supersymmetry,  the eight-dimensional cone over $X_7$ with metric
\be
\label{cone}
ds^2=dr^2+r^2 ds^2(X_7)
\ee
must have special holonomy. If the cone has $Spin(7)$ holonomy then $X_7$ is a weak $G_2$
manifold and the $D=11$ solution is dual to an $\caln=1$ SCFT. If the
cone has $SU(4)$ holonomy, i.e. it is a Calabi-Yau four-fold, then
$X_7$ is a Sasaki-Einstein manifold and this is dual to an $\caln=2$
SCFT. If the cone has $Sp(2)$ holonomy, i.e. is hyper-K\"ahler,
then $X_7$ is tri-Sasakian and this is dual to an $\caln=3$ SCFT.
If the cone is flat we have the maximally supersymmetric
$AdS_7\times S^7$ solution dual to an $\caln=8$ SCFT.
This is summarised in the table below.
Note that for
SCFTs in three-dimensions with $\caln$ supersymmetries the
$R$-symmetry is $so(\caln)$.
\begin{center}
\begin{tabular}{|l|l|l|}
\hline
$X_7$ &${\rm Holonomy\,\, of\,\,} C(X_7)$ &  $\caln$
\cr
\hline
Weak $G_2$ & $Spin(7)$ &  1
\cr
Sasaki-Einstein & $SU(4)$ (Calabi--Yau) & 2
\cr
tri-Sasaki & $Sp(2)$ (hyper-K\"ahler) &  3
\cr
$S^7$ & $\id$ (flat) &  8\cr
\hline
\end{tabular}
\end{center}

One class of examples that we study is when $X_7$ is homogeneous
i.e. of the form $G/H$. The classification of all homogeneous
Einstein manifolds in seven dimensions was carried out long ago in
\cite{Castellani:1983yg}
(see \cite{Duff:1986hr} for a review) and discussed in an $AdS$/CFT setting in \cite{Castellani:1998nz}.
We have:
\begin{enumerate}

\item
the Sasaki-Einstein space\footnote{$M(3,2)$ is also known as $M^{1,1,1}$.}
$M(3,2)$ with $SU(3)\times SU(2)\times U(1)$ isometry \cite{Witten:1981me,Castellani:1983mf}.
This is a regular Sasaki-Einstein manifold that is naturally a $U(1)$ bundle
over $\bCP^2\times S^2$.

\item
the Sasaki-Einstein space $Q(1,1,1)$ with $SU(2)^3\times U(1)$
isometry \cite{D'Auria:1983vy}.
This is also a regular Sasaki-Einstein manifold that is naturally a $U(1)$ bundle over
$S^2\times S^2\times S^2$

\item
the Aloff-Wallach spaces $N(k,l)=SU(3)/U(1)$ (see
\cite{Cvetic:2001zx} for a recent discussion). For general $k,l$
these admit two Einstein metrics, $N(k,l)_I$ and $N(k,l)_{II}$ but
if $l=0$ these two metric coincide \cite{Page:1984ac}.
The metric\footnote{$N(1,1)_I$ is also known as $N^{0,1,0}$.}
$N(1,1)_I$ is tri-Sasakian \cite{Castellani:1983mf} while its squashed version
$N(1,1)_{II}$ as well as all other metrics on $N(k,l)$ are weak $G_2$ \cite{Castellani:1983mf,Page:1984ac}.
Note that both metrics on $N(1,1)$ have $SU(3)\times SO(3)$ isometry,
while the other metrics on $N(k,l)$ have $SU(3)\times U(1)$ isometry.

\item
the squashed seven sphere, $SO(5)\times SO(3)/SO(3)\times SO(3)$. This manifold is
weak $G_2$ and has $SO(5)\times SO(3)$ isometry.

\end{enumerate}
Note that the weak $G_2$ manifold $SO(5)/SO(3)_{\rm max}$ does not
have a $T^3$ action so that we cannot deform it using the method
of
\cite{Lunin:2005jy}. The Sasaki-Einstein homogeneous space
$V_{5,2}=SO(5)\times U(1)/SO(3)\times U(1)$ does have a $T^3$
action but it does not preserve any Killing-spinors and hence the
deformed solution will necessarily break supersymmetry.

In addition we will also deform the $AdS_4\times X_7$ solutions
where $X_7$ is the new infinite class of Sasaki-Einstein spaces
found in \cite{Gauntlett:2004hh}. These spaces are co-homogeneity
one generalisations of the $M(3,2)$ and $Q(1,1,1)$ spaces. In
order to simplify the presentation, we will not explicitly discuss
the generalisation discussed in
\cite{Gauntlett:2004hs,Chen:2004nq}, nor the further generalisation
discussed in \cite{Cvetic:2005ft}, but it is straightforward to
do so.

A summary of the supersymmetric solutions that we consider
(including the $S^7$ example studied in \cite{Lunin:2005jy}) and
the amount of supersymmetry that we will show the deformed
solutions preserve, ${\cal N}_\gamma$, is summarised below:
\be
\begin{tabular}{|l|c|c|l|}
\hline
$X_7$ & ${\cal N}$ & ${\cal N}_\gamma$ & Examples
\cr
\hline
$S^7$ & 8 & 2 &
\cr
Tri-Sasaki & 3 & 1 & $N(1,1)_I$
\cr
Sasaki-Einstein & 2 & 2 & $Q(1,1,1), M(3,2)$ and\, generalisations.
\cr
Weak $G_2$ & 1 & 1 & Squashed $S^7$, $N(k,l)_{I,II}$\cr
\hline
\end{tabular}
\ee

\subsection{Field theory}
\label{fieldtheory}

\begin{figure}
\centerline{\epsfxsize=5cm\epsfbox{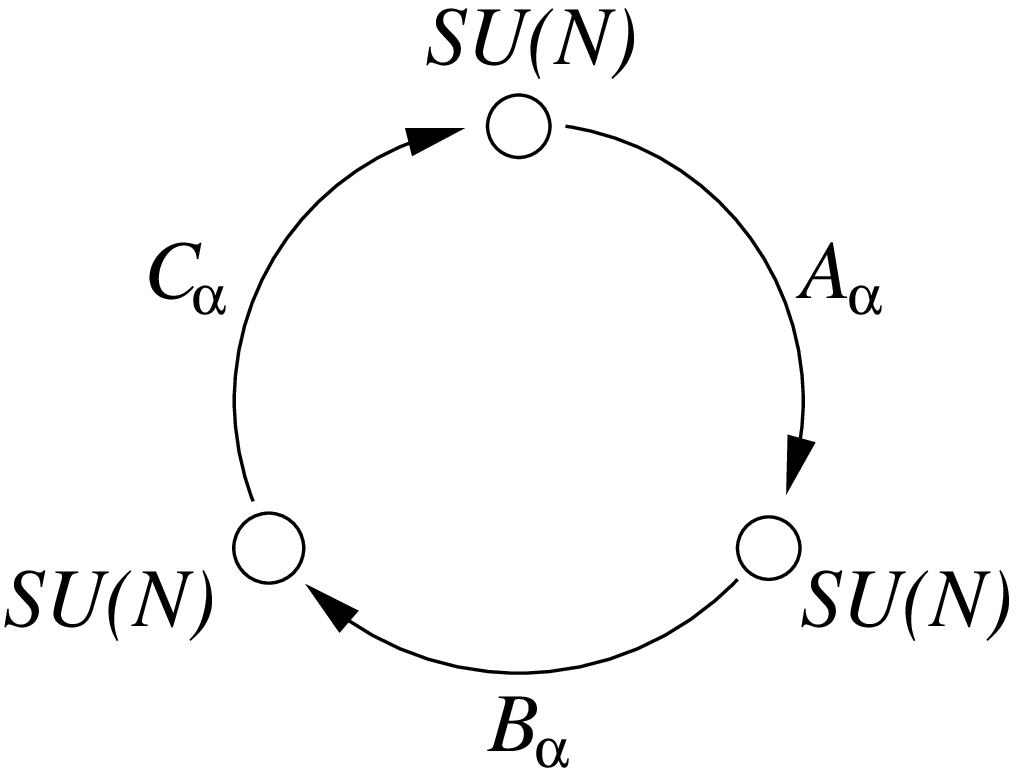}\hskip 2cm\epsfxsize=6cm\epsfbox{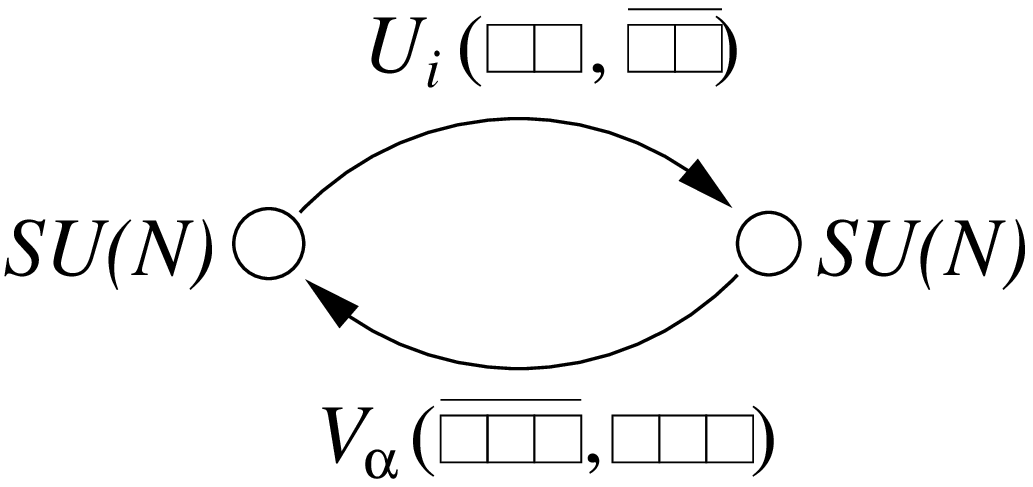}}
\caption{The quiver diagrams for the $Q(1,1,1)$ and $M(3,2)$ field theories. \label{QMquiver}}
\end{figure}

\begin{figure}
\centerline{\epsfxsize=6cm\epsfbox{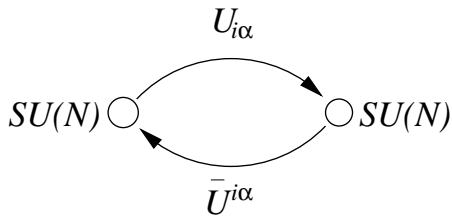}}
\caption{The quiver diagram for the $N(1,1)_I$ field
  theory. \label{N010quiver}}
\end{figure}

The above $AdS_4\times X_7$ solutions are dual to the
$d=3$ supersymmetric CFTs arising on M2-branes sitting at the apex of
the singular cone. Although our understanding of such field
theories is still quite rudimentary, some concrete proposals have
been made for the cases of  $M(3,2)$ and  $Q(1,1,1)$ in
\cite{Fabbri:1999hw} and  $N(1,1)_I$ in \cite{Billo:2000zr} (see
also \cite{Gukov:1999ya}) and some remarkable tests have been
performed. The field theories arise from the IR limit of quiver
gauge theories whose field content is summarized in Fig.
\ref{QMquiver} and Fig. \ref{N010quiver}.

The $\caln=2$ theory dual to $Q(1,1,1)$ theory has gauge group
$SU(N)^3$. The isometries of $Q(1,1,1)$ give rise to
$SU(2)^3\times U(1)$ global symmetry, where the $U(1)$ factor is
the $R$-symmetry. There are three chiral fields $A$, $B$ and $C$
which transform under the gauge group and $SU(2)^3$ in the
representations summarised in the table below:
\be
\begin{tabular}{|c|c|c|c|c|c|c|}
\hline            & $SU(N)_1$ & $SU(N)_2$ & $SU(N)_3$ & $SU(2)_1$& $SU(2)_2$& $SU(2)_3$\cr\hline
$A_\a$      &   $\rep{N}$     &    $\rep{\bar N}$      &   $\rep{1}$     &  $\rep{2}$ & $\rep{1}$&$\rep{1}$\cr
$B_\a$   &  $\rep{1}$     &   $\rep{N}$     &   $\rep{\bar{N}}$     & $\rep{1}$&$\rep{2}$ &$\rep{1}$\cr
$C_\a$   &  $\rep{\bar N}$     &    $\rep{1}$     &  $\rep{N}$    & $\rep{1}$&$\rep{1}$ &$\rep{2}$\cr
\hline
\end{tabular}
\ee
They each have conformal dimension 1/3 (equal to their $U(1)_R$
charge). The chiral operators, which agree with the Kaluza-Klein
spectrum, are given by $\Tr(ABC)^k$ symmetrised over all $SU(2)$
indices, i.e. in the $(\rep{k+1},\rep{k+1},\rep{k+1})$ rep, and have conformal
dimension $k$.
Note that one has to assume that other $SU(2)$ reps decouple in
the IR (unlike the case of $T^{1,1}$ studied in
\cite{Klebanov:1998hh} there is no superpotential suitable for
this task). The second betti number of $Q(1,1,1)$,
$b_2(Q(1,1,1))$, is two so the AdS/CFT correspondence predicts
that the field theory has an extra $U(1)^2$ baryonic global
symmetry. The manifold $Q(1,1,1)$ has three supersymmetric
5-cycles (corresponding to divisors in the eight-dimensional cone)
and wrapping five-branes on these cycles predicts baryons with
conformal dimension $N/3$ in accord with the operators $\det A$,
$\det B$ and $\det C$. Furthermore, for the global symmetry
groups, the representations of the five-brane states and the
operators have also been shown to agree.

Let us next describe the $\caln=2$ theory dual to $M(3,2)$,
which has gauge group $SU(N)^2$. The isometries
of $M(3,2)$ give rise to $SU(3)\times SU(2)\times U(1)$ global symmetry, where the $U(1)$ factor is again
the $R$-symmetry. There are two chiral fields $U$, and $V$ which transform in the following representations:
\be
\begin{tabular}{|c|c|c|c|c|}
\hline            & $SU(N)_1$ & $SU(N)_2$ & $SU(3)$& $SU(2)$\cr\hline
$U_i$      &  $Sym^2(C^N)$     &   $Sym^2(C^{N*})$    &   $\rep{3}$     &  $\rep{1}$ \cr
$V_\a$   &  $Sym^3(C^{N*})$     &   $Sym^3(C^N)$    &   $\rep{1}$     & $\rep{2}$\cr
\hline
\end{tabular}
\ee
The field $U$ has conformal dimension 4/9 while $V$ has dimension
1/3. The chiral operators given by ${\rm Tr}(U^3V^2)^k$, totally
symmetrised over the $SU(3)\times SU(2)$ indices, have dimension $2k$
and these agree with the Kaluza-Klein spectrum. Again one has to
assume that other $SU(3)\times SU(2)$ reps decouple in the IR.
Since $b_2(M(3,2))=1$ the field theory is predicted to have a
$U(1)$ baryonic symmetry. The manifold $M(3,2)$ has two
supersymmetric 5-cycles and wrapping five-branes on these cycles
predicts baryons with conformal dimensions $4N/9$ and $N/3$ in
accord with the operators $\det U$ and $\det V$, respectively. In
addition, for the global symmetry groups, the representations of
the five-brane states and the operators have also been shown to
agree.

The $\caln=3$ field theory dual to $N(1,1)_I$ has gauge group
$SU(N)^2$. The isometries of $N(1,1)_I$ are (locally) $SU(3)\times
SU(2)$ where the $SU(2)$ factor is the $R$-symmetry. The theory
contains three hypermultiplets transforming as a triplet under
$\SU(3)$. In terms of
$\caln=2$ superfields, these can be written as two sets of chiral
fields $u^i$ and $v_i$, $i=1,2,3$ transforming in the $\rep{3}$ and
the $\rep{\bar{3}}$ of $SU(3)$ respectively. The $SU(2)_R$ action is
best described by grouping these fields into
$U^i_{\a}=(u^i,-\bar{v}^i)$ and $V_{i \a}=(v_i,\bar{u}_i)=
-\epsilon_{\a\b} \bar{U}_i^{\b}$, each transforming as a doublet of
$SU(2)_R$.  We can summarise this is as:
\be
\begin{tabular}{|c|c|c|c|c|}
\hline            & $SU(3)$ & $SU(2)$ & $SU(N)_1$ & $SU(N)_2$ \cr\hline
$U^i_\a$    &   $\rep{3}$     &    $\rep{2}$      &   $\rep{N}$     &  $\rep{\bar{N}}$ \cr
$V_{i\a}$   &  $\rep{\bar{3}}$     &    $\rep{2}$      &   $\rep{\bar{N}}$     &  $\rep{N}$ \cr
\hline
\end{tabular}
\ee
Unlike in the other two field theories discussed above, it was
proposed in~\cite{Billo:2000zr} (using an $\caln=2$ superfield
description) that one can write down a suitable superpotential which
does not vanish after taking the trace over the gauge indices.
The resulting equations of motion imply $u^iv_ju^j=-u^jv_ju^i$ and
$v_iu^jv_j=-v_ju^jv_i$. It would be interesting to formulate this
directly in an $\caln=3$ language.

The chiral operators with dimension $k$ that agree with the
Kaluza-Klein spectrum can be written as
\be
{\rm Tr} \,\, U^{(i_1} V_{(j_1}\dots  U^{i_k)} V_{j_k)}
\ee
where it must be assumed that the operators are symmetrised over $SU(3)$ indices
and $SU(2)$ indices (not shown). We also demand that the operators are
actually traceless and hence in the irreducible representation $(\rep{k},\rep{k})$ of $SU(3)$ corresponding
to the Young tableau:
\begin{eqnarray}
&&
\begin{array}{l}
\begin{array}{|c|c|c|c|c|c|}
\hline
             \hskip .3 cm & \cdots & \hskip .3 cm &
             \hskip .3 cm & \cdots & \hskip .3 cm \nn
\hline
\end{array}\\
\begin{array}{|c|c|c|}
             \hskip .3 cm & \cdots & \hskip .3 cm \nn
\hline
\end{array}
\end{array}\ \\
&& \,\,\,
\underbrace{\hskip 2.2 cm}_{k}
\underbrace{\hskip 2.2 cm}_{k}
\label{su3young}
\end{eqnarray}
Other reps are assumed to decouple in the IR. It is interesting to
note that the superpotential ensures that some of these
operators are traceless. The theory has a single $U(1)$ baryonic
symmetry since $b_2(N(1,1)_I)=1$. We show in appendix~\ref{baryon}
that $N(1,1)_I$ has supersymmetric 5-cycles whose volumes give rise
to baryons with dimension $N/2$, in agreement with the conformal
dimension of the operator $\det U$.

Although we shall not be considering it much in the following we
note that a field theory dual to the $V_{5,2}$ case is discussed
in \cite{Ceresole:1999zg}. As far as we are aware no proposal has
been made for the case of the squashed seven sphere nor for other
$N(k,l)_{I,II}$ cases. The field theories dual to the new
Sasaki-Einstein spaces presented in \cite{Gauntlett:2004hh} have not
yet been worked out, but it seems plausible that once the toric
description of the corresponding Calabi-Yau fourfold cones is found,
generalising the results of \cite{Martelli:2004wu}, a proposal can be
found by following the arguments of \cite{Fabbri:1999hw}.

\section{Deformation of Supergravity Solutions}
\label{deform}

In this section, after introducing our conventions for $D=11$
supergravity, we will present the generating procedure of
\cite{Lunin:2005jy} in a form convenient for our purposes. We also
analyse how much supersymmetry can be preserved in the deformed
solutions.

\subsection{Conventions}

Our conventions for the bosonic fields of $11d$ supergravity are as
in~\cite{Gauntlett:2002fz}. In particular the Lagrangian is
\be
2\kappa^2 {\cal L} = 
R *1 - \undos F_4 \wedge * F_4 - \frac{1}{6} C_3 \wedge F_4
\wedge F_4 \,.
\ee
The $11d$ Planck length is defined by
$2 \kappa^2 = (2\pi)^8 l_{11}^9$.
In these conventions, the dimensionless integer valued M2- and M5-brane
charges are expressed in terms of the flux of $F_4$ as
\be
\label{Mcharge}
N_{M2} = \frac{1}{(2\pi l_{11})^6} \int_{C_7} * F_4
\sac
N_{M5} = \frac{1}{(2\pi l_{11})^3} \int_{C_4} F_4 \, ,
\ee
for some 7-cycle $C_7$ or 4-cycle $C_4$ surrounding the branes.

The main subject of this paper is deformations of backgrounds of
the type $AdS_4 \times X_7$ which arise as the near horizon limit of branes
sitting at the apex of the eight-dimensional cone with base $X_7$.
The undeformed solution takes the form
\be
\label{undeformed}
ds^2 = R_{X_7}^2 \left( \frac{1}{4} ds^2_{AdS_4} +
ds^2_{X_7}\right),
\;\;\; F_{(4)} = \frac{3}{8}  R_{X_7}^3 vol_{AdS_4} \, ,
\ee
provided that we normalize $ds^2_{X_7}$ such that $R_{\m\n}(X_7) = 6
g_{\m\n}(X_7)$, just as for a unit radius $S^7$.
The radius $R_{X_7}$ is
determined by the quantisation condition \bref{Mcharge},
\be \label{y7radius}
6 R_{X_7}^6 \mbox{Vol}(X_7) = (2\pi l_{11})^6 N \,,
\ee
with Vol$(X_7)$ being the volume of $X_7$. For the special case of
a unit radius $S^7$ we have Vol$(S^7) = \pi^4/3$ and hence
$R_{S^7} = (32\pi^2 N )^{1/6} l_{11}$.
Finally, we note that
the choice of orientation for $AdS_4$ (or equivalently $X_7$)
implicit in the expression for the four-form is chosen so that the solution
is supersymmetric: recall that if, for example, we flip the sign of $F_4$
then we obtain a solution to the equations of motion that does not
preserve any supersymmetry (apart from the special case when $X_7$ is
the round seven-sphere)~\cite{Duff:1983nu}.

\subsection{Transformation Rule}
\label{sec-transf-rule}

The full symmetry group of the $11d$ supergravity compactified on
a three-torus is $SL(3,\CR) \times SL(2,\CR)$. We will consider
supergravity solutions with $U(1)^3$ isometry and deform them by
an element of $SL(2,\CR)$. The action of the symmetry group can be
deduced from the following Kaluza-Klein ansatz:\footnote{
To avoid clutter, we will often suppress the overall radius
$R$ in the intermediate steps.
}
\bea \label{ansatz}
ds_{11}^2
&=& \Delta^{1/3} M_{ab} D\vph^a D\vph^b + \Delta^{-1/6} g_{\m\n}dx^\m dx^\n \,,
\\ \label{ansatz2}
C_3 &=& C_{(0)} D\vph^1 D\vph^2 D\vph^3 +
\half C_{(1)ab} D\vph^a D\vph^b + C_{(2)a} D\vph^a  + C_{(3)} \,,
\eea
where $D\vph^a = d\vph^a + A_{\m}^a dx^\m$, $\det(M_{ab})=1$ and the products
of forms in the last line are wedge products.
The $SL(3,\CR)$ part of the symmetry is manifest. The
transformation rule under the $SL(2,\CR)$ part is less trivial and
can be found, for example, in \cite{8dsugra}. The $8d$ Einstein
metric $g_{\m\n}$ and $C_{(2)a}$ are invariant. The
$\tau$-parameter is defined by
$\tau \equiv -C_{(0)} + i \sqrt{G}_{T^3} =  -C_{(0)} + i \Delta^{1/2}$
and it transforms in the usual way
\bea\label{usualtran}
\L = \begin{pmatrix} a & b \cr c & d \end{pmatrix} \;\;\; \in SL(2,\CR) ~; \;\;\;\;
\tau \goto \frac{a \tau + b }{ c \tau +d} \,.
\eea
The $8d$ vectors $A^a_\m$ and $C_{(1)ab\m}$ form a doublet in the
same way as the NSNS and RR two-form fields do in the $10d$ IIB
supergravity,
\bea \label{pot}
B^a = \begin{pmatrix} A^a \cr -\half \e^{abc} C_{(1)bc}
\end{pmatrix}
\sac
B^a \goto \L^{-T} B^a \,.
\eea
Finally, and somewhat unexpectedly, the field strength corresponding to
$C_{(3)}$ also forms a doublet with its magnetic dual:
\bea
\label{f4}
H = \begin{pmatrix} F_{(4)} \cr \Delta^{1/2} *_8 F_{(4)} + C_{(0)}
F_{(4)}\end{pmatrix} \,,
\;\;\;\;\;
H \goto \L^{-T} H \,,
\eea
where the Hodge dual is taken with respect to the $8d$ metric.

Now, we are ready to apply the transformation rules to deform the
background~\bref{undeformed} using a $U(1)^3$ action on $X_7$.
Since the 4-form field strength lies entirely in $AdS_4$, we have
$C_{(0)}=C_{(1)}=C_{(2)}=0$ in the Kaluza-Klein
ansatz~\eqref{ansatz2}, and the only non-vanishing field strength is
$F_{(4)}=dC_{(3)}$. As shown in~\cite{Lunin:2005jy} the deformed
metric is regular only for the $SL(2,\CR)$ elements of the form
\be\label{stel}
\L = \begin{pmatrix} 1 & 0 \cr \g & 1\end{pmatrix} \,.
\ee
We shall only consider these transformations and
call the corresponding transformed solutions ``$\gamma$-deformed solutions''.

It follows that once we express the metric in the KK form (\ref{ansatz}),
$g_{\m\n}$, $A^a$ are invariant, while
$C_{(1)ab}$, $C_{(2)a}$ remain zero. There are only two
non-trivial transformation rules. The first is
$\tau' = \tau/(1+\g \tau)$ which gives
\be
\label{defG}
\Delta' = G^2 \Delta
\sac
C_{(0)}' = -\g G \Delta \,,
\ee
where
\be
G\equiv\frac{1}{(1+\gamma^2\Delta)} \,.
\ee
Combining this with the other transformation concerning  $F_{(4)}$, we find
\be
F_4' = F_4 - \g \Delta^{1/2} *_8 F_4 -
\g d \left( G \Delta D\vph^1 D\vph^2 D\vph^3\right).
\ee

To summarise, to obtain the $\gamma$-deformed solution we first
write the metric on $X_7$ in the form
\be
\label{reorg}
ds^2_{X_7} = ds^2_{T^3} + ds^2_4 \sac ds^2_{T^3} :=
\Delta^{1/3} \, M_{ab} D\vph^a D\vph^b \, ,
\ee
with $D\vph^a\equiv d\vph^a+A^a$ and $\det(M_{ab})=1$. The quantities $\Delta$, $M_{ab}$ and $A^a$ will depend on
the particular $X_7$ being considered. The solution obtained by the $SL(2,\CR)$ transformation
(\ref{stel}) is simply
\bea \label{final-conf-1}
ds^2_{11} &=& G^{-1/3} \left( \frac{1}{4} ds^2_{AdS_4}  \,\,+\,\,
ds^2_4 \,\,+ \,\,G \,\, ds^2_{T^3} \right) \,,
\\ \label{final-conf-2}
F_{4} &=&
\frac{3}{8} \, vol_{AdS_4}\,\, -\,\, 6 \,{\g}\, \Delta^{1/2} \,vol_{ds_4}
- \g \, d(G\Delta \,\, D\vph_1 \we D\vph_2 \we D\vph_3) \, ,
\eea
where
$vol_{ds_4}$ is the volume form\footnote{The orientation for $ds_4$ is chosen so that
the positive orientation for $X_7$, canonically fixed
by it's Killing spinors, is equal to $vol_{ds_4}d\varphi^1 d\varphi^2 d\varphi^3$.}
of $ds_4^2$
and $G= (1+ \g^2 \Delta)^{-1}$.
The dependence on the radius $R_{X_7}$ can be reinstated by multiplying $ds_{11}^2$
by
$R_{X_7}^2$, $F_4$ by $R_{X_7}^3$ and replacing $\g$ by $\hat{\g} =
(R_{X_7}/l_{11})^3 \g$.

It is interesting to consider the term $6\g \Delta^{1/2} vol_{ds_4}$
in the four-form field strength. In the case that the base-space $M_4$,
with local metric $ds^2_4$, is actually a manifold, this term corresponds
to $\g N$ M5-branes wrapped on the $T^3$, where $N$ is, of course,
the number of membranes creating the background. To see this, we first re-instate
$R_{X_7}$ dependence which brings this piece of the 4-form, which we denote by $f_4$,
to the form
\be
f_4 = 6 \frac{R_{X_7}^6}{l_{11}^3}  \g \Delta^{1/2} vol_{ds_4} \,.
\ee
Using \bref{Mcharge} and \bref{y7radius}, we then have
\be\label{myers5}
N_{M5} = \frac{1}{(2\pi l_{11})^3} \int_{M_4} f_4 =
\frac{\g N}{ \mbox{Vol}(X_7)} \int_{M_4} (2\pi)^3 \Delta^{1/2}
vol_{ds_4}=\g N .
\ee
We have used the fact that if the co-ordinates $\vph^a$ have period $2\pi$, then $(2\pi)^3\Delta^{1/2}$ is the volume of the $T^3$ fibre at any
point on the base $M_4$ and the second integral then equals Vol$(X_7)$.
These M5-branes are sitting at a constant radial position in
$AdS_4$ and they wrap the $T^3$. The latter is a contractible
cycle so this configuration can be interpreted as
membranes expanded to a form a fuzzy $T^3$ via the Myer's effect.

We note that choosing the coordinates $\vph^a$ to have period $2\pi$ ensures that the
$SL(2,\bZ)$ transformations with integer entries in (\ref{usualtran}) is the exact $SL(2,\bZ)$ symmetry
of the full M-theory. This is consistent with $N_{M5}$ in (\ref{myers5}) being integer valued when $\gamma$ is.

\subsection{Preservation of Supersymmetry}
\label{susy}

We focus on deformations which preserve some supersymmetry.
Thus we demand that the $T^3$ action should commute with at least one supersymmetry.
i.e.  the Lie derivative with respect to the three $U(1)$'s of at least one Killing spinor should vanish.
If this is not the case, then the deformed solution will not preserve any supersymmetry
and we briefly return to this possibility at the end of the paper.

We normalise our seven-dimensional metrics so that $Ricci(X_7)=6 g(X_7)$. Choosing the
seven-dimensional hermitian gamma-matrices to be imaginary and satisfy $\{\gamma_a,\gamma_b \}=2\delta_{ab}$,
the real Killing spinors $\psi$ satisfy
\be
\nabla_a\psi=\pm \frac{i}{2}\gamma_a\psi.
\ee
The maximally supersymmetric
case of $S^7$ has eight Killing spinors for each sign, but for all other cases the Killing spinors have the same sign (which
is fixed by the orientation of $X_7$). Take it to be the plus sign for definiteness.
If $X_7$ admits a Killing vector $k^a$ then we can define the spinorial Lie-derivative. Acting on a Killing spinor
we have
\bea\label{expkil}
\CL_k\psi&\equiv&[k^a\nabla_a+\frac{1}{4}\nabla_ak_b\gamma^{ab}]\psi\nn
&=&[\frac{i}{2}k^a\gamma_a+\frac{1}{4}\nabla_ak_b\gamma^{ab}]\psi \,.
\eea
It is well known that the spinorial Lie-derivative preserves the space of Killing spinors (see e.g. \cite{Figueroa-O'Farrill:1999va}).

Consider now $X_7$ to be a weak $G_2$ manifold with precisely one Killing spinor $\psi$.
These give solutions dual to $\caln=1$ CFTs which have no $R$-symmetry.
If $k$ is a Killing vector on $X_7$ then $\CL_k\psi$ is also a Killing spinor and
hence proportional to $\psi$:  $\CL_k\psi=\alpha\psi$ with $\alpha$
real. However, multiplying by $\psi^T$ and using the fact that $\gamma^a$ and
$\gamma^{ab}$ are both anti-symmetric we conclude that $\alpha=0$. Thus the
single Killing spinor must be invariant. Thus, quite generally, any weak $G_2$ manifold with
a $T^3$ action can be deformed and the deformed solution will preserve one
supersymmetry. We will discuss some explicit examples below.

Next consider $X_7$ to be Sasaki-Einstein with two Killing spinors.
These give solutions dual to $\caln=2$ CFTs which have a $U(1)$ $R$-symmetry (or, in fact, a non-compact $\bR$ $R$-symmetry),
and this corresponds to the fact that the Sasaki-Einstein manifold has a canonical Killing-vector,
sometimes called the Reeb vector. Identifying $U(1)$ with $SO(2)$, the two Killing spinors transform as a $\rep{2}$
(see e.g. \cite{Figueroa-O'Farrill:1999va}). Thus if we are seeking deformed solutions that preserve supersymmetry
we cannot use the Reeb vector as part of the $T^3$ action. It is not difficult to show, using (\ref{expkil}),
that if a $T^3$ action leaves one of the two Killing spinors invariant, it will in fact leave both invariant.
To see this, we can take $\psi_1$ and $\psi_2$ be two orthogonal Killing spinors with $\CL_k\psi_1=0$ and
$\CL_k\psi_2=\alpha\psi_1+\beta\psi_2$. If we now multiply the second equation by $\psi_2^T$ we deduce that $\beta=0$. On the other
hand, if we multiply by $\psi_1^T$ and use the first equation we then deduce that $\alpha=0$.
Solutions deformed using such a $T^3$ action will thus preserve two supersymmetries.

Finally, we consider $X_7$ to be tri-Sasaki with three Killing spinors.
These give solutions dual to $\caln=3$ CFTs which have an $SO(3)$ $R$-symmetry,
and this corresponds to the fact that the Sasaki-Einstein manifold has three canonical Killing-vectors
generating an $SO(3)$. The three Killing spinors transform as a $\rep{3}$ of $SO(3)$
(see e.g. \cite{Figueroa-O'Farrill:1999va}).
The tri-Sasaki-Einstein manifolds and $T^3$ actions that we consider lead to deformations that preserve one supersymmetry.
We first choose a $U(1)$ sub-group of the $SO(3)$ $R$-symmetry.
Two of the Killing spinors are charged with respect to this $U(1)$ while the third
is neutral. Thus, a suitable $T^3$ action is obtained by supplementing this $U(1)$ action with a $T^2$ action
that also leaves this single Killing spinor invariant.

We now turn to some detailed examples.

\section{Examples}

In this section we construct the $\gamma$-deformed solutions for a number of
different examples. As explained above, for each example, we just need to
compute the quantities ($\Delta,A^i,M_{ab},ds^2_4$)
defined in \bref{reorg} for a specified $T^3$ action.
The metric and the four-form of the $\gamma$-deformed solution is then given by
\bref{final-conf-1} and \bref{final-conf-2}.

\subsection{Sasaki-Einstein}

Let us begin by recalling a few facts about Sasaki-Einstein (SE) manifolds.
Locally, any seven-dimensional SE metric can be put in the form
\be
ds^2=(d\psi'+\sigma)^2+ds^2_6(M_6) \,,
\ee
where $ds^2_6$ is a local metric on a K\"ahler-Einstein base-space $M_6$. If
$J_6$ and $\Omega_6$ are the K\"ahler-form and $(3,0)$-form on $M_6$ respectively, we have
$d\sigma=2J_6$ and $d\Omega_6=4i\sigma\wedge \Omega_6$.
The vector $\partial_{\psi'}$ is the Reeb vector which is Killing.
If the Reeb vector has closed orbits and the $U(1)$ action is free, we have
a regular SE manifold. In this case $M_6$ is a manifold.
If the $U(1)$ action has finite isotropy groups then we have a quasi-regular SE and
$M_6$ is an orbifold. If the orbits of the Reeb vector do not close we have
an irregular SE, and $M_6$ is not globally defined in any sense. The field theories dual
to regular and quasi-regular SE manifolds have a $U(1)$ $R$-symmetry, while those dual to
irregular SE manifolds have a non-compact $\bR$ $R$-symmetry.

We seek SE manifolds with a $T^3$ action commuting with the two Killing spinors.
The two Killing spinors can be constructed locally from the gauge-covariantly constant spinors
on the K\"ahler-Einstein base space (see e.g. \cite{hart}). From this description we deduce that the $T^3$ action must be independent
of the Reeb vector and leave the gauge-covariantly constant spinors on the base invariant.
A convenient way to check the latter is to check that the $T^3$ action leaves $J_6$ and $\Omega_6$ invariant.

We will consider the homogeneous examples $M(3,2)$ and $Q(1,1,1)$, both of which are regular SE manifolds.
$M(3,2)$ is a $U(1)$ bundle over $\bCP^2\times S^2$ with winding numbers 3 and 2 over $\bCP^2$ and $S^2$, respectively.
$Q(1,1,1)$ is a $U(1)$ bundle over $S^2\times S^2\times S^2$ with unit winding over each $S^2$.
These metrics are incorporated in the cohomegeneity one manifolds found in \cite{Gauntlett:2004hh} and we will
consider the entire family. This family includes both quasi-regular and irregular SE manifolds.

We will follow the description of the relevant $X_7$ given in Ref.
\cite{Gauntlett:2004hh}, where a general method of building up a SE
manifold $Y_{2n+3}$ from a K\"ahler-Einstein manifold $B_{2n}$ is
explained. For the case at hand, $n=2$, the metric is given by
\cite{Gauntlett:2004hh}
\bea \label{SEgeneral}
ds^2_{X_7} = U(\r)^{-1} d\r^2 + \r^2  ds^2_{B_4} + q(\r) \left(
d\psi+ j_1\right)^2 + \w(\r) \left( d\a + f(\r) (d\psi+j_1)
\right)^2 \, ,
\eea
where $ds^2_{B_4}$ is the metric of the $4d$ base and the one-form
$j_1$ satisfies $dj_1 = 2 J$, with $J$ being the K\"ahler form of
$B_4$. The normalisation condition $R_{\m\n}(X_7) = 6 g_{\m\n}(X_7)$
is equivalent to $R_{\m\n}(B_4) = 2 g_{\m\n}(B_4)$. In this
convention, the functions of $\r$ appearing in (\ref{SEgeneral})
are given by
\bea
U(\r)&=& \frac{1}{3} - \r^2 + \frac{\kappa}{768 \r^6}
\sac
\w(\r) = \r^2 U(\r) + (\r^2 -1/4)^2 \,,
\nn
f(\rho) &=& \frac{ \r^2\,(U(\r) + \r^2 - 1/4) }{ \w(\rho) }
\sac
q(\r) = \frac{\r^2 U(\r)}{ 16 \,\,\w(\r)} \, .
\eea
See \cite{Gauntlett:2004hh} for more discussion on the possible
values that $\kappa$ can take and the range of the co-ordinate $\rho$.
The two possible examples of $B_4$ with explicitly known metric
are $\CC P^1 \times \CC P^1$ and $\CC P^2$. We will discuss both
of them.

The isometry group of $X_7$ always includes a $U(1)^4$ factor. Two
of the $U(1)$'s come from the base $B_4$. With a suitable choice
of coordinates, the two $U(1)$'s will be generated by
$(\partial_{\phi_1},\partial_{\phi_1})$. The other two $U(1)$'s
are $(\partial_{\alpha}, \partial_{\psi})$ appearing in
\bref{SEgeneral}. As explained in \cite{Gauntlett:2004hh}, the combination
$V = \partial_{\psi} - \partial_{\a}$ corresponds to the
Reeb vector or equivalently the $R$-symmetry of the superconformal algebra.
We will perform the $SL(2,\CR)$ transformation on the $T_3$
generated by $(\pa_{\phi_1},\pa_{\phi_2},\pa_{\alpha})$. In each case we have
checked that the action does indeed leave the two Killing spinors invariant.
We also note, that in general, the coordinate $\alpha$ does not have period $2\pi$.

\subsubsection{$S^2\times S^2$ base}

If we choose the four dimensional base $B_4=S^2 \times S^2$, then the metric on $X_7$ is
\bea \label{SE}
ds^2_7 &=& U^{-1} d\r^2 + \frac{\r^2}{2} \left( d\t_1^2 +
\sin^2
\t_1 d\phi_1^2 + d\t_2^2 + \sin^2 \t_2   d\phi_2^2\right) \nn
&& + q \left( d\psi+ j_1\right)^2 + \w \left( d\a + f (d\psi+j_1) \right)^2 \,,
\eea
where the 1-form $j_1 =-\cos \t_1 d\phi_1 -\cos \t_2 d\phi_2$ is
such that $2 J_2 = dj_1$, $J_2$ being the K\"ahler form $B_4$.
We have chosen the radius squared of the two $S^2$s
to be 1/2 to give the correct normalisation.

As explained above, to obtain the $\gamma$-deformed solution
corresponding to the $T^3$ generated by
$(\pa_{\phi_1},\pa_{\phi_2},\pa_\a)$, it suffices to give:
\bea
\Delta &=& {\r^2 \w \over 4} \left[ q(1-c_{2\t_1} c_{2\t_2})+ \r^2 s^2_{\t_1}s^2_{\t_2} \right] \,, \nn
A^1 &=& -{q \r^2 \w c_{\t_1} s_{\t_2}^2 \over 2 \Delta} d\psi
\sac
A^2 =-{q \r^2 \w c_{\t_2} s_{\t_1}^2 \over 2 \Delta} d\psi
\sac
A^3 ={f \r^4 \w s_{\t_1}^2 s_{\t_2}^2 \over 4 \Delta} d\psi \,,
\nn \nn
M_{ab} &=&\Delta^{-1/3} \,\,
\begin{pmatrix} (q+\w f^2) c_{\t_1}^2 + \undos \r^2 s_{\t_1}^2  & (q+\w f^2) c_{\t_1}c_{\t_2} & -f \w c_{\t_1}  \cr
\cdot & (q+\w f^2) c_{\t_2}^2 + \undos \r^2 s_{\t_2}^2 &-f \w c_{\t_2}  \cr
\cdot&\cdot& \w \end{pmatrix} \,, \nn
ds^2_4 &=& U^{-1} d\r^2 + {\r^2\over 2} \left( d\t_1^2 +  d\t_2^2
\right) + {q \w
\r^4 s_{\t_1}^2 s_{\t_2}^2 \over 4 \Delta} d\psi^2 \, , \label{m4-1} \nonumber
\eea
where we have introduced the notation $s_\theta=\sin\theta$, $c_\theta\equiv \cos\theta$, etc.
Note also that we have not entered all elements of the symmetric matrix $M$.
The $D=11$ solution is then given by \bref{final-conf-1} and \bref{final-conf-2}.

\subsubsection{$\bCP^2$ base}

We use the following metric for $\bCP^2$:
\be
{ds^2_{\CC P^2} \over 3} = \sum_{i=1}^3 d\m_i^2 + \m_1^2d\phi_1^2 + \m_2^2d\phi_2^2
-(\m_1^2d\phi_1+\m_2^2d\phi_2)^2
\sac
\sum_{i=1}^3 \m_i^2 =1 \,.
\ee
We have chosen the $\bCP^2$ to have radius squared 3 to give the correct
normalisation. With this choice of the radius, the
K\"ahler form and potential read
\bea
J_2 &=& 3\left( \m_1 d\m_1 d\phi_1 + \m_2 d\m_2 d\phi_2 \right) \,,
\nn j_1 &=& 3\left( \m_1^2 d\phi_1 + \m_2^2 d\phi_2 \right) \sac
2J_2 = dj_1 \,.
\eea
These coordinates are inherited from the well known reduction
$\CC^3\goto S^5 \goto \CC P^2$,
\be
w_1 = \m_1 e^{i(\psi + \phi_1)},
\;\;
w_2 = \m_2 e^{i(\psi + \phi_2)},
\;\;
w_3 = \m_3 e^{i\psi} \,,
\ee
and are related to the standard inhomogeneous coordinates with the
Fubini-Study K\"ahler potential $K$ in the obvious way,
\bea
K = 3\log(1+|z_1|^2 +|z_2|^2) \sac  z_i = w_i/w_3
\sac i=1,2 \,.
\eea
The 7-manifold is
\be \label{SEcp2}
ds^2_7 = U^{-1} d\r^2 + {\r^2}  ds^2_{\CC P^2} + q \left( d\psi-
j_1\right)^2 + \w \left( d\a + f (d\psi-j_1) \right)^2 \,.
\ee

The computation of the relevant data for the $\gamma$-deformed solution
using the $T_3$ generated by $(\pa_{\phi_1},\pa_{\phi_2},\pa_\a)$ give:
\bea
\Delta &=& 9 \r^2 \w  \mu_1^2 \mu_2^2 \left[ 3 q (\mu_1^2+\mu_2^2) + \r^2 \mu_3^2) \right] \,,\nn
A^1 &=& {9 q \r^2 \w \mu_1^2 \mu_2^2 \over \Delta} d\psi
\sac
A^2 ={9 q \r^2 \w \mu_1^2 \mu_2^2 \over  \Delta} d\psi
\sac
A^3 ={9 f  \r^4 \w \mu_1^2 \mu_2^2 \mu_3^2 \over \Delta} d\psi \,,
\nn \nn
\Delta^{1/3} M_{ab} &=&
\begin{pmatrix} 9 (q+\w f^2) \mu_1^4 + 3 \r^2 \mu_1^2 (1-\mu_1^2)  & 3 (3 q+ 3 \w f^2-\r^2) \mu_1^2 \mu_2^2 & 3 f \w \mu_1^2  \cr
\cdot &9(q+\w f^2) \mu_2^4 + 3 \r^2 \mu_2^2 (1-\mu_2^2)   &  3 f \w \mu_2^2\cr
\cdot&\cdot& \w \end{pmatrix}
\nn
ds^2_4 &=& U^{-1}
d\r^2 + {3 \r^2}\sum_i d\mu_i^2 + { 9 q w \r^4  \mu_1^2\mu_2^2 \mu_3^2 \over
\Delta} d\psi^2 \,. \nonumber
\eea

\subsection{Tri-Sasaki}
\label{N11}

We next consider the tri-Sasaki metric on $N(1,1)$. Topologically
$N(1,1)$ is the coset space $SU(3)/U(1)$ where the $U(1)$ is embedded
in the $\l_8=\frac{1}{\sqrt{3}}{\rm diag}(1,1,-2)$ direction.
As a bundle, one way it can be viewed is a single $\SO(3)$ instanton
on $\bCP^2$
(the analogue of viewing $S^7$ as an instanton on $S^4$).
It admits two Einstein metrics: $N(1,1)_I$ is tri-Sasaki, while
$N(1,1)_{II}$ is weak $G_2$. In coordinates adapted to the $\SO(3)$
fibration, both metrics can be written in the form~\cite{Page:1984ac},
\bea
2 ds^2 &=&
d\mu^2+\frac{1}{4}\sin^2\mu(\sigma_1^2+\sigma_2^2)+\frac{1}{4}\sin^2\mu
\cos^2\mu \,\sigma_3^2 \label{metpp}
\\
&& +\ \lambda^2
\left\{(\Sigma_1-\cos\mu\sigma_1)^2
+(\Sigma_2-\cos\mu\sigma_2)^2
+(\Sigma_3-{\textstyle \frac{1}{2}}(1+\cos^2\mu)\sigma_3)^2
\right\} \nonumber \,,
\eea
where $\Sigma_i$ are right invariant one-forms on $\SO(3)$, and
$\sigma_i$ are right invariant one-forms on $\SU(2)$. This form of the
metric is discussed in appendix~\ref{baryon}. The parameter
$\lambda^2=1/2$ for $N(1,1)_I$ and $\lambda^2=1/10$ for
$N(1,1)_{II}$. The isometry group for both $N(1,1)_I$ and
$N(1,1)_{II}$ is $SU(3)\times SO(3)$.

For the purpose of deforming the $N(1,1)_I$ metric and also
comparing to the corresponding deformation of the field theory, it
is convenient to use an alternative set of coordinates, naturally
adapted to the construction of the metric as a hyper-K\"ahler
quotient (see e.g. ~\cite{Billo:2000zr}). The relation between the
two set of coordinates is explained in appendix~\ref{coord}. The
quotient construction starts with the flat metric for $\CC^6$
parametrised by
$(u^i,v_i)$. There is an action of $\SU(3)$ with $u^i$ transforming as
in the $\rep{3}$ representation and $v_i$ in the $\rep{\bar{3}}$
representation. There is also an $\SU(2)$ action where
$(u^i,-\bar{v}^i)$ transforms as a doublet. These actions descend to
$N(1,1)_I$ and generate the $\SU(3)\times\SO(3)$ isometry group. Note
that the representations are the same as those of the fields in the dual
field theory described in section~\ref{fieldtheory} and hence we use the
same notation. The hyper-K\"ahler quotient is built from the
$U(1)$ action $(u^i,v_i)\to(e^{i\theta} u^i, e^{-i\theta}
v_i)$. Viewing $\CC^6$ as three copies of the quaternions, this action
preserves the corresponding triplet of complex structures. The
resulting moment maps give the constraints
\begin{equation}
   |u^i|^2 - |v_i|^2 = 0 = u^i v_i .
\end{equation}
If, in addition, we mod out by $(u^i,v_i)\sim(e^{i\theta} u^i,
e^{-i\theta}v_i)$, the resulting space is the eight-dimensional cone
over $N(1,1)_I$. Fixing the radial direction by taking
$|u^i|^2=|v_i|^2=1$ gives the metric on $N(1,1)_I$. As discussed in
appendix~\ref{coord}, before modding out by the $U(1)$ action the
constrained $(u^i,v_i)$ satisfying $|u^i|^2=|v_i|^2=1$ and
$u^iv_i=0$ actually parametrise $\SU(3)$. In this way one can see
that
$N(1,1)_I\simeq\SU(3)/U(1)$.

To deform $N(1,1)_I$ we need to pick out three $U(1)$ Killing
directions: a $U(1)^2$ from the $\SU(3)$ part of the isometry group
and a $U(1)$ from the $\SO(3)$ part. Since the $\SO(3)$ isometry
descends from the $\SU(2)$ action on $(u^i,-\bar{v}^i)$ writing
\begin{equation}
\label{wzdef}
   u^i = e^{i\theta+i\phi_3/2}w^i , \qquad
   v_i = e^{-i\theta+i\phi_3/2}z_i ,
\end{equation}
we see that $\theta$ parametrises the $U(1)$ action we mod out by
and
$\phi_3$ parametrise the $U(1)$ subgroup of $\SU(2)$. (Note that the
$\SU(2)$ element $\diag(-1,-1)$ is actually part of the $U(1)$ factor
by which we mod out and hence the $\SU(2)$ action descends to an
$\SO(3)$ action on $N(1,1)_I$.) The constraints $|u^i|^2=|v_i|^2=1$
imply that $w^i$ and $z_i$ parametrise a pair of $\bCP^2$
manifolds. In addition $u^iv_i=0$ implies $w^iz_i=0$. In terms of
these constrained variables, reducing from the flat metric on
$\CC^6$ it is easy to show that the metric on $N(1,1)_I$ has the
form
\be
\label{metuv2}
   2 \, ds^2 = d\bar{w}_i dw^i - |\bar{w}_i dw^i|^2
      +  d\bar{z}^i dz_i-|\bar{z}^i dz_i|^2
      + \tfrac{1}{2}(d\phi_3 - i(\bar{w}_i dw^i
         + \bar{z}^i dz_i))^2 ,
\ee
The term $d\bar{w}_i dw^i - |\bar{w}_i dw^i|^2$ is the standard
Fubini--Study metric of $\bCP^2$ and similarly for $z_i$, though
one must remember, of course that the coordinates are related by
the constraint $w^iz_i=0$. We can introduce an explicit
parametrisation of $w^i$ and $z_i$,
\begin{align}
   (w^1,w^2,w^3) &=
      e^{-i(\a_1+\a_2)/3}(\m_1 e^{i\a_1},\m_2e^{i\a_2},\m_3) \,,
      && \m_1^2 + \m_2^2 +\m_3^2 = 1\,, \nn
   (z_1,z_2,z_3) &=
      e^{-i(\b_1+\b_2)/3}(\n_1 e^{i\b_1},\n_2 e^{i\b_2},\n_3) \,,
      && \n_1^2 + \n_2^2 +\n_3^2 = 1 \,.
\end{align}
The component parts of the metric can then be written as
%
\begin{align}
   d\bar{w}_i dw^i - |\bar{w}_i dw^i|^2
      &= \sum_{i=1}^{3} d\m_i^2
         + \mu_1^2(1-\mu_1^2)d\alpha_1^2
         + \mu_2^2(1-\mu_2^2)d\alpha_2^2
         - 2\mu_1^2\mu_2^2d\a_1d\a_2,\nn
   -i\bar{w}_i dw^i
      &= (\mu_1^2-\frac{1}{3})d\alpha_1+(\mu_2^2-\frac{1}{3})d\alpha_2\,,
\label{metuv3}
\end{align}
%
with similar expressions associated to $z_i$. The constraint $w^iz_i=0$
then reads
\begin{equation}
   e^{i(\a_1+\b_1)}\m_1\n_1 + e^{i(\a_2+\b_2)}\m_2\n_2
      + \m_3\n_3 = 0 .
\end{equation}

Finally we must identify the $U(1)^2$ subgroup of the $\SU(3)$
isometries. Changing variables
\begin{equation}
   \a_r = \rho_r + \phi_r, \qquad
   \b_r = \rho_r - \phi_r, \qquad r=1,2 \,,
\end{equation}
it is easy to see that the $U(1)^2$ subgroup acts by shifting
$\phi_r$ leaving $\rho_r$ invariant. In addition the constraint
$w^iz_i=0$ now involves only $\rho_r$, namely
\begin{equation}
\label{wzcond}
   e^{i2\rho_1}\m_1\n_1 + e^{i2\rho_2}\m_2\n_2 + \m_3\n_3 = 0 .
\end{equation}
In summary, $\phi_1$, $\phi_2$ and $\phi_3$ parametrise the
maximal
$U(1)^3$ of the $\SU(3)\times\SU(2)$ isometry group which we use for
the $\gamma$-deformation. Since there is no real three dimensional
representation of $\SU(3)$, the three Killing spinors on $N(1,1)_I$
must be invariant under the action of $\SU(3)$ though transform as a
triplet under the $\SO(3)$ R-symmetry. Following the discussion in
section~\ref{susy}, we thus conclude that this $U(1)^3$ action
preserves a single Killing spinor.

Following the general discussions in the section~\ref{deform}, we
can now give the data required to obtain the $\gamma$-deformed solution.
%
%
We find
\begin{align}
   16\Delta &= \m_1^2\m_2^2\m_3^2 + \n_1^2\n_2^2\n_3^2
      + \left[\m_1^2\n_1^2 (\m_2^2\n_3^2+\m_3^2\n_2^2)
         + ({\rm cyclic})\right]\, , \nn
   \Delta^{1/3} M_{ab} &=
      \frac{1}{4}\begin{pmatrix}
         -(\mu_1^2+\nu_1^2)(\mu_1^2+\nu_1^2-2)&
            -(\mu_1^2+\nu^2_1)(\mu_2^2+\nu_2^2)&
            (\mu_1^2-\nu_1^2)  \cr
         \cdot & -(\mu_2^2+\nu_2^2)(\mu_2^2+\nu_2^2-2) &
            (\mu_2^2-\nu^2_2)  \cr
         \cdot&\cdot& 1
      \end{pmatrix}\nn
   16 \Delta A^1 &=
      \left[ \mu_1^2\mu_2^2\mu_3^2
         + \mu_1^2\nu_2^2 (1-\mu_1^2-\nu_2^2 +\mu_1^2 \nu_2^2 )
         - (\mu \leftrightarrow \nu) \right] d\rho_1 \nn
      &\qquad
      + 2 \left[ \mu_1^2\mu_2^2\nu_2^2 (\nu_2^2-1)
         - (\mu \leftrightarrow \nu) \right] d\rho_2 \, , \nn
   16 \Delta A^2 &=
      \left[ \mu_1^2\mu_2^2\mu_3^2
         - \mu_1^2\nu_2^2 (1-\mu_1^2-\nu_2^2 +\mu_1^2 \nu_2^2 )
         - (\mu \leftrightarrow \nu) \right] d\rho_2 \nn
      &\qquad
      + 2 \left[ \mu_1^2\mu_2^2\nu_1^2 (\nu_1^2-1)
         - (\mu \leftrightarrow \nu) \right] d\rho_1 \, , \nn
   24 \Delta A^3 &=
      \left[- \mu_1^2\mu_2^2\mu_3^2
         - \mu_1^2\nu_2^2 (1-\mu_1^2-\nu_2^2 +\mu_1^2 \nu_2^2 )
         + 3 \mu_1^2\mu_2^2\nu_1^2 (1+\mu_3^2 -\nu_1^2) \right. \nn
      &\qquad \left. -2 \mu_1^2 \mu_2^2\nu_1^2\nu_2^2
         + (\mu \leftrightarrow \nu) \right]  d\rho_1
      + \left[ 1 \leftrightarrow 2 \right] d\rho_2  \,, \nn
   ds^2_4 &= {\undos} \sum_i^3 (d\mu_i^2 + d\nu_i^2)
      + ds^2_{\rho_1,\rho_2} \,,\nn
   -8 \Delta \, ds^2_{\rho_1,\rho_2}
      &\equiv \left[ \mu_1^2 \mu_2^2 \mu_3^2 \nu_1^2(\nu_1^2-1)
         + (\mu \leftrightarrow \nu) \right] d\rho_1^2
      + \left[ 1 \leftrightarrow 2 \right] d\rho_2^2 \nn
      &\qquad
      + 2 \mu_1^2\mu_2^2\nu_1^2\nu_2^2(\mu_3^2+\nu_3^2) d\rho_1 d\rho_2
      \label{hi1} \,.
\end{align}

In the discussion of the gauge theory dual, it will be useful to know
the action of the Weyl group of $\SU(3)$ on the deformation. In
terms of elements of $\SU(3)$ one is interested in the discrete subgroup
generated by the maps
\begin{align}
\label{Weyl}
   a_1 : (u^1,u^2,u^3) \mapsto (u^2,-u^1,u^3), \nn
   a_2 : (u^1,u^2,u^3) \mapsto (u^1,u^3,-u^2), \nn
   a_3 : (u^1,u^2,u^3) \mapsto (-u^3,u^2,u^1),
\end{align}
together with similar expressions for $v_i$. The Weyl group is then
the quotient of this subgroup by those elements in Cartan subgroup
$U(1)^2\subset\SU(3)$. The original metric is clearly invariant under
this group since it is a subgroup of the isometry group $\SU(3)$. The
question is whether the deformed metric is also invariant. The action
on our coordinates for the first element $a_1$ in~\eqref{Weyl} is
straightforward: it permutes the pairs
$(\mu_1,\mu_2)$, $(\nu_1,\nu_2)$, $(d\phi_1,d\phi_2)$ and
$(d\rho_1,d\rho_2)$. The other elements are a little more
complicated. For instance the second element $a_2$ translates into
\be
\begin{pmatrix}\m_1' \cr \m_2'\end{pmatrix}
= \begin{pmatrix}\m_1 \cr (1-\m_1^2-\m_2^2)^{1/2}\end{pmatrix},
\;\;\;\;
\begin{pmatrix}d\phi_1' \cr d\phi_2'\end{pmatrix}
=
\begin{pmatrix}1 & -1 \cr 0 & -1\end{pmatrix} \begin{pmatrix}
d\phi_1 \cr d\phi_2\end{pmatrix}
\equiv S \begin{pmatrix}d\phi_1 \cr d\phi_2\end{pmatrix} \,,
\ee
with similar expressions for $\n_r$ and $\rho_r$.
It is straightforward to see that this transformation leaves the
deformed metric invariant whereas it changes the sign of the part of
$F_4$ linear in $\g$. We return to this issue in
section~\ref{sec-deform}.

\subsection{Weak $G_2$}

We argued above that any weak $G_2$ manifold, with precisely one
Killing spinor, admitting an isometric $T^3$ action will lead to a deformed
solution preserving one supersymmetry. One example in this class is
$N(1,1)_{II}$, the squashed Einstein metric on $N(1,1)$.  It is
straightforward to construct the deformed solution using the
coordinates in \bref{metpp}, but as the expressions are a bit
complicated we will not present them here. More general Aloff-Wallach
spaces $N(k,l)$ have two weak $G_2$ metrics each with $SU(3)\times
U(1)$ isometry and as this includes a $T^3$ action we can similarly
construct the $\gamma$-deformed solutions.

The weak $G_2$ example that we will explicitly consider here is the squashed seven-sphere. This is given by the
coset $Sp(2)\times Sp(1)/Sp(1)\times Sp(1)$ and has $SO(5)\times SO(3)$ isometry
\cite{Awada:1982pk}. The squashing can be obtained by viewing the $S^7$ as an $S^3$ bundle
over $S^4$ and changing the size of the $S^3$ fibres relative to the $S^4$ base. Apart from the
round $S^7$ metric, this procedure leads to only one more Einstein metric given by
\bea \label{squashed}
{20 \over9} \, ds^2_{squashed} &=& d\mu^2 + {1 \over 4} \sin^2 \mu \, \s_i^2 + {1 \over 5} (\Sigma_i-\cos^2{\mu \over 2} \,\, \s_i)^2 \,,
\eea
where the $\s$'s and $\Sigma$'s are left-invariant one forms in $SU(2)$.
Here we take
\bea \s_1&=&\cos\psi d\theta+\sin\psi\sin\theta d\phi \,,\nn
\s_2&=&- \sin\psi d\theta+\cos \psi\sin\theta d\phi \,, \nn
\s_3&=&d\psi+\cos\theta d\phi  \,.\eea
The same expressions hold for the $\Sigma$'s, putting `primes' over the angles.
Note also that the factor of ${20/9}$ ensures that  $R_{\mu\nu}=6 g_{\mu\nu} $.

In these co-ordinates, only an $SU(2)\times SU(2) \times U(1)$ isometry is manifest,
where the $U(1)$ corresponds to simultaneous shifts of $\psi$ and $\psi'$.
In order to perform the deformation, we define $\psi_{\pm} =(\psi \pm \psi')/4$
and select the $T^3$ generated by the Killing vectors
$(\partial_{\phi},\partial_{\phi'},\partial_{\psi_+})$. With the definitions given, the three angles
on $T^3$ have the same period $2\pi$.

As explained above, the deformation is completely specified by computing the following data:
\footnotesize
\bea
\a^2 \Delta &=& -16 \ct^2 \sm^4 - 16(3+2 \cm) \cmh^4 (\ct \ctp+ \cps \st \stp)^2  \smh^2 \nn
&&- 32 \ct \ctp \cmh^2 (\ct \ctp+ \cps \st \stp) \sm^2 \smh^2 +
(4 \cmh^4 + 5 \sm^2) (5 \sm^2 + 4 \stp^2 \smh^4) \,, \nn \nn
{\a^2 \over 2} \Delta A^1 &=& 2 \sm^2(\stp ( 5 \ct (\ctp \cmh^2  \sps  d\t +4  \stp \, d\psi_-)+
\cps \cmh^2 \st (-20 \ctp d\psi_- + (3+2 \cm)  \sps \stp \, d\t)) \nn
&&+ \sps \st (-3-2 \cm+ \ctp^2 \smh^2) \, d\t' \, ) \,,
\nn
{\a^2 \over 2} \Delta A^2 &=& 2 \cmh^2 \sm^2 (-5 \ctp \st(\ct  \sps d\t' +4 (3-2 \cm)  \st \, d\psi_-)+
\stp (\cps \st (20 \ct d\psi_- \nn
&&  -(3+2 \cm)  \sps \st \, d\t' )+ \sps (5+16 \cmh^2 \st^2 \smh^2) \,d\t \,)) \,,
\nn
{\a^2 \over 2} \Delta A^3 &=& \sm^2 (\cmh^4 (- \sps \stp (\ctp+4 \ct^2 \ctp +2 \cps s_{2\t}  \stp) \, d\t
+  \st^2 (4+\ctp^2-4 \cps^2 \stp^2) \, d\psi_-) \nn
&&+ 5  \smh^2 (2+3 \ctp^2-(4+\ctp^2) \smh^2) \, d\psi_- + \cmh^2 ( \sps \st (\ct (4+\ctp^2)
+\cps \ctp \st \stp) d\t'  \nn
&&- 5 \ctp \sps \stp \smh^2 \, d\t
 +  (2-12 \ct^2+3 \ctp^2 +7 \ct^2 \ctp^2-2 \cps^2 \cm \st^2 \stp^2 \nn
 &&-(-4+9 \ct^2) (4 + \ctp^2) \smh^2)  \, d\psi_- )) \,,
\nn \nn
\b^2 M_{ab} &=&\Delta^{-1/3} \,\,
\begin{pmatrix}
4 \cmh^4 + 5 \sm^2  &  -4\cmh^2 (\ct \ctp+\cps \st \stp) & 2\ct (-4 \cmh^2+4 \cmh^4+ 5 \sm^2)
 \cr
\cdot  &4 & 8 \ctp \smh^2
 \cr
\cdot&\cdot&   4 (5 \sm^2 +4 \smh^4)
\end{pmatrix} \,, \nonumber
\eea
\normalsize
where $\a \equiv 2000/27$ and $\b\equiv 20/3$. The expression for $ds^2_4$ is
a bit long, so we prefer to leave it as implicitly defined by
\be
ds^2_4 = ds^2_{squashed} - \Delta^{1/3} M_{ab} (d\vp^a+A^a) (d\vp^b+A^b)  \,.
\ee
Again, after the deformation, the resulting configuration just amounts to substituting
these values in \bref{final-conf-1}-\bref{final-conf-2}.

\section{Deformations of the dual field theory}
\label{sec-deform}

The $\gamma$-deformed supergravity solutions we studied in the
previous section should correspond to an exactly marginal
deformation of the dual $d=3$ conformal field theory. We will
now identify the relevant operators for the cases in which
a proposal for the dual field theory has been made.

Lunin and Maldacena showed that for the $d=4$ CFTs with $AdS_5$ duals,
the $\gamma$-deformation results in a sort of star-product amongst the
elementary fields. This has the simple effect of introducing various
$\gamma$-dependent phases into the Lagrangian. This result was deduced
using open string field theory and the fact that the
$\gamma$-deformation is a result of a T-duality, a shift and another
T-duality. This argument cannot be directly applied to the $d=3$ CFTs
studied here, because there is no analog of deriving the CFTs from
open strings.

However, for small values of $\gamma$, there is an alternative way to find
the exactly marginal operator. At the linearised level, the
deformation in supergravity amounts to turning on a massless
($\Delta = d$) mode in the KK spectrum of the original background.
The AdS/CFT dictionary then gives the exactly marginal operator in the
CFT dual, which is the corresponding change in the dual field theory
Lagrangian.

Consider the deformations of ${\cal N}=4$ SYM studied in \cite{Lunin:2005jy},
for example, which preserve $\caln=1$ supersymmetry. The deformation in the
field theory must correspond to perturbing the superpotential.
The only term in the deformed supergravity solution that is linear in
$\gamma$ appears in the combination $B_{NS}+i B_{RR}$, which translates
into an operator obtained by acting $Q_\a Q^\a$ on the chiral
primary operator ${\rm Tr}(\Phi^1 \Phi^2 \Phi^3 + \Phi^3
\Phi^2 \Phi^1)$. Since the superpotentials enters the Lagrangian in the
form $(\int d^2\theta W + c.c.)$, we conclude that the $\gamma$-deformation
turns on the superpotential $W = {\rm Tr}(\Phi^1 \Phi^2 \Phi^3 + \Phi^3 \Phi^2 \Phi^1)$.
In fact an easier way to obtain this superpotential is to observe that
it is the unique  $\Delta=3$ chiral primary operator which breaks the
$SU(3)$ global symmetry to $U(1)^2$. For small $\gamma$ it indeed agrees with
the general result of \cite{Lunin:2005jy}: $W={\rm Tr}(e^{i\pi\gamma}\Phi^1 \Phi^2 \Phi^3 -
e^{-i\pi\gamma} \Phi^1 \Phi^3 \Phi^2)$.

Essentially the same story holds for all
the examples considered in \cite{Lunin:2005jy} and those that we consider here.
Starting with a theory preserving 4 supercharges, the
$\gamma$-deformation amounts to adding a superpotential to the field
theory Lagrangian
which (a) has $\Delta = d - 1$, (b) belongs to a short-multiplet in the
superconformal algebra, and (c) breaks the global symmetry group
into $U(1)^2$, for AdS$_5 \times M_5$ examples, or $U(1)^3$ for AdS$_4
\times X_7$ examples. This observation provides a quick way to identify
the superpotential even without looking at the detailed form of
the $\gamma$-deformed supergravity solution. For example, in the
$T^{1,1}$ theory, $W = {\rm Tr}(A_+ B_+ A_- B_- + A_+ B_- A_- B_+)$
is the unique $\Delta=3$ chiral primary operator which breaks the
$SU(2)^2$ global symmetry to $U(1)^2$, in agreement, for small $\gamma$, with
the general result of \cite{Lunin:2005jy}:
$W = {\rm Tr}(e^{i\pi\gamma}A_+ B_+ A_- B_- -e^{-i\pi\gamma} A_+ B_- A_- B_+)$.

Let us now turn to the $d=3$ examples for which the dual CFT is
known, as reviewed in section 2.

\vspace{4 mm}
\noindent
{\bf Q(1,1,1) and M(3,2) cases}
\vspace{3 mm}

The deformed solutions for
$Q(1,1,1)$ and $M(3,2)$ both preserve $\caln=2$ supersymmetry, so we seek
a superpotential that is chiral primary with dimension two and preserves
$U(1)^2$. For the $Q(1,1,1)$ theory, we note that the chiral operators
${\rm Tr}(ABC)^2$, totally symmetrised over the $SU(2)$ indices, have
$\Delta=2$ and transform in the $(\rep{3},\rep{3},\rep{3})$ of
$SU(2)^3$. This contains a single operator that is neutral under
$U(1)^3$ and hence this must be the superpotential giving the
$\gamma$-deformation.

A similar story holds for the $M(3,2)$ theory. The operator
${\rm Tr}(U^3V^2)$ in the totally symmetrized product
$(\rep{10},\rep{3})$ of $SU(3)\times SU(2)$, is chiral with $\Delta=2$. This
again contains a unique state neutral under $U(1)^3$ and hence this must be
the superpotential giving the $\gamma$-deformation.

\vspace{4 mm}
\noindent
{\bf N(1,1)$_I$ case}
\vspace{3 mm}

Our last example is the $N(1,1)_I$ theory. From our discussion in
section 4 we know that the $\gamma$-deformation breaks $\caln=3$ to
$\caln=1$ and $SU(3)\times SO(3)_R$ to $U(1)^3$. Note that despite not
having any surviving $R$-symmetry, there is a $U(1)\subset SO(3)_R$
which still acts on the matter fields and leads to a global symmetry
of the action.

This breaking implies that the deformation of the SCFT must be by a
term which preserves only $\caln=1$ supersymmetry. As we reviewed in
subsection~\ref{fieldtheory}, the theory contains three $\caln=3$
hypermultiplets which in $\caln=2$ language correspond to two chiral
superfields $u^i$ and $v_i$ which can be naturally grouped as
$U^i_\alpha=(u^i,-\bar{v}^i)$ and the conjugate
$V_{i\alpha}=(v_i,\bar{u}_i)$. We can parametrise $\caln=3$
superspace by supercoordinates $\theta^\pm, \theta^0$, which under
$U(1)\subset SO(3)$ carry charge $\pm1,0$ respectively. The chiral
fields $u^i$ and $\bar{v}^i$ are then expansions in terms of
$\theta^+$ and $\theta^-$ respectively. However, we can also represent
the hypermultiplets in terms of $\caln=1$ multiplets, that is, as
superfields which are functions of the $\caln=1$ superspace with
supercoordinate $\theta^0$. In both descriptions, we can use the
same name for a superfield and its lowest scalar component.
However, since we are keeping a different subset of supercharges
from $\caln=3$, the scalars are combined with different components
of fermions to form the superfields.

In section~\ref{fieldtheory} we described the chiral primary
operators in the conformal field theory using an $\caln=2$
language. In particular, the most general chiral primary operator of
dimension two is of the form
\be
T_{ij}^{kl} \,\, C^{\a\b\g\d} \,\,
   {\rm Tr} \,\, U^i_\a\, V_{j\b}\, U^k_\g\, V_{l\d} \,,
\ee
where $C^{\a\b\g\d}$ and $T_{ij}^{kl}$ are irreducible tensors in the
$(\rep{27},\rep{5})$ of $SU(3)\times SU(2)_R$. Given the comments of
the previous paragraph, we now view this operator as being written
in terms of $\caln=1$ superfields $U$ and $V$. If we now integrate
over the full $\caln=1$ superspace we get a marginal deformation
that breaks $\caln=3$ to $\caln=1$. Next, we need to recall that
the
$\gamma$-deformed solution is invariant under $U(1)\subset
SO(3)_R$. But only one element in the $\rep{5}$ representation has this
property: this implies that the only non-zero component of $C$, up to
symmetrization, is $C^{1122}$. Finally, invariance under
$U(1)^2\subset SU(3)$ still leaves three possibilities,
as there are three states in the $\rep{27}$ of $SU(3)$ invariant under
the Cartan subgroup.

In order to decide which of these three operators is turned on in the
$\gamma$-deformed supergravity solution~\bref{hi1}--\bref{final-conf-2},
we consider the transformation properties under the discrete
subgroup generated by the elements~\eqref{Weyl}. Acting on the
$\rep{27}$ representation, this
subgroup transforms the three $U(1)^2$-invariant states into each
other. In fact, it gives a realization of the Weyl group of
$\SU(3)$. Furthermore, there is a unique combination of states which is
invariant under all elements of the Weyl group. Expanding the
$\gamma$-deformed solution in the parameter $\gamma$, to leading order,
the metric is unchanged, but there is a deformation of the flux $F_4$
as given in~\eqref{final-conf-2} which, according to the Kaluza--Klein
analysis, should transform as the $\rep{27}$ of $\SU(3)$. Under the
cyclic subgroup of the Weyl group generated by the element $b=a_2a_1$
it is easy to see that the flux deformation is invariant. Thus it would
appear that the deformation corresponds to the unique Weyl-singlet
combination of operators. To complete this analysis one must show that
the flux is also invariant under $a_i$ separately. As stands, it
appears that under these elements the linear term in $F_4$ maps to
minus itself. Since there are no elements in the $\rep{27}$
representation with such properties this raises a puzzle.

\section{Discussion}

We have used the techniques of \cite{Lunin:2005jy} to construct a
large number of new supersymmetric solutions of $D=11$ supergravity
with $AdS_4$ factors. They correspond to exactly marginal
deformations in the dual CFT. For certain examples, we managed
to identify the operators in the CFT giving rise to the deformation for
small values of the deformation parameter $\gamma$.

It would be interesting to explore this further. In particular it
would be nice to be able to identify the exact deformation in the
field theory for all values of $\gamma$ as was done in the string theory
setting in \cite{Lunin:2005jy}. It would also be interesting to know
if the field theories considered here have additional exactly marginal
deformations and if so whether they too can be constructed in explicit form.

We have focused on deformations of supersymmetric
$AdS_4$ solutions that preserve some supersymmetry. We now briefly
discuss how some simple generalisations lead to new $AdS_4$
solutions with no supersymmetry. Of course the stability of such
solutions, and hence their relevance to the AdS/CFT
correspondence, now needs to be checked.

One possibility is to start with a supersymmetric solution and
choose a $T^3$ action that doesn't preserve any supersymmetry.
There are several possibilities of this type for the solutions considered in
this paper. In addition one could also consider the supersymmetric
solution based on $V_{5,2}$ which has a $T^3$ action but not one
commuting with supersymmetry.

Another possibility is to deform a non-supersymmetric $AdS_4$ solution with a $T^3$ action.
One case is to start with the supersymmetric solutions of this paper and simply reverse the
sign of the four-form (or equivalently, reverse the orientation on $X_7$).
As is well known this gives solutions that do not preserve
any supersymmetry \cite{Duff:1983nu} but are nevertheless stable
\cite{Duff:1984sv}. The $\gamma$-deformed solution has metric as in
\bref{final-conf-1} and the four-form as in \bref{final-conf-2} but with
the sign of the first two terms on the right hand side changed.

A second case is to recall that some of the supersymmetric
solutions considered in this paper are members of more general
families of homogeneous Einstein spaces. For example, $M(3,2)$ is
a member of the $M(m,n)$ family which are all $U(1)$ bundles over
$\bCP^2\times S^2$ (this is the same family as the $M^{pqr}\sim
SU(3)\times SU(2)\times U(1)/SU(2)\times U(1)\times U(1)$ spaces
introduced in \cite{Witten:1981me}). Similarly, $Q(1,1,1)$ is a
member of the $Q(p,q,r)$ family which are all $U(1)$ bundles over
$S^2\times S^2\times S^2$ (this is the same family as the
$Q^{p,q,r}\sim SU(2)^3\times U(1)/U(1)^3$ introduced in
\cite{D'Auria:1983vy}). These give rise to stable non-supersymmetric
solutions as discussed in \cite{Page:1984ad} and \cite{Page:1984ae}
for the $M(m,n)$ and $Q(p,q,r)$ families, respectively. The
$\gamma$-deformed versions of these can easily be found.

A third case to consider is the non-supersymmetric ``Englert ''
solutions \cite{Englert:1982vs,Duff:1982ev,Awada:1982pk}.
These are obtained from a supersymmetric solution by switching on
a four-form flux constructed as a bi-linear of a Killing spinor
and flipping the orientation. It is known that they do not preserve any
supersymmetry \cite{Englert:1983qe,Biran:1986zg}.
Furthermore, they have been shown to be unstable if the
supersymmetric solution preserves two or more supersymmetries
\cite{Page:1984fu}. In addition the Englert solution using the
squashed $S^7$ was demonstrated to be unstable in
\cite{Ito:1984uj}, while some of the $N(k,l)$ have shown to be
unstable
\cite{Page:1984fu,Yasuda:1984vt}. If we assume that there are some
stable $N(k,l)$ Englert solutions then we can generate additional
stable solutions. We need to check that the four-form is invariant
under the $T^3$ action (in order to be able to reduce the solution
to $D=8$ and then employ the generating procedure), but this is
guaranteed by its construction as a Killing spinor bilinear and,
as noted at the beginning of section 4, the fact that the Killing spinor is
invariant under the $T^3$ action.

A final possibility, again with non-vanishing four-form, are the
solutions based on $U(1)$ bundles over $D=6$ K\"ahler spaces
\cite{Pope:1984bd,Pope:1984jj}. We are unaware of
any results concerning the stability of this class of solutions.
Nevertheless, some of these solutions admit a $T^3$ action
and we can generate the corresponding $\gamma$-deformed solutions.

\bigskip\noindent{\bf Note added:}
The results of this paper were presented by one of us (JPG) at the
meeting {\it Workshop on Gravitational Aspects of String Theory} May
2-6, 2005 at Fields Institute, Toronto. Following this, but as we were
finalising this preprint, the preprint \cite{AV} appeared which has
some overlap with our results.

\appendix

\section*{Appendix}

\section{Baryons from $N(1,1)_I$}
\label{baryon}

Baryonic states in field theory are dual to five-branes wrapping
five-cycles in the manifold $X_7$. For supersymmetric states the
five cycles must lift to supersymmetric six-cycles in the
eight-dimensional cone $C(X_7)$ given in~\eqref{cone}. In the case
of $N(1,1)_I$, the cone is hyper-K\"ahler and so has $\Symp(2)$
holonomy. A supersymmetric six-cycle on $C(N(1,1)_I)$ can at most
be calibrated by the Hodge dual of one of the three K\"ahler
forms. Equivalently, it will be holomorphic (a divisor) with
respect to one of the three complex structures. As such it breaks
one of the three supersymmetries on $C(N(1,1)_I)$. Consequently,
in looking for supersymmetric cycles it is natural to concentrate
on only one of the complex structures on the cone. Equivalently,
one isolates one of the three Sasaki structures on $N(1,1)_I$.

Viewed as a regular Sasakian manifold, $N(1,1)_I$ can be written
canonically as an explicit $U(1)$ fibration over a K\"ahler--Einstein
(KE) six-manifold:
\be\label{canform}
ds^2=(d\psi'+\sigma)+ ds^2_6 \,.
\ee
In fact $ds^2_6$ is the KE metric on the flag manifold
$SU(3)/U(1)\times U(1)$, normalised so that $\Ricci_6=8g_6$
and $d\sigma=2J_6$. To make this structure explicit, we start with
the metric in the standard form of an $\SO(3)$ fibration over
$\bCP^2$, as given in~\cite{Page:1984ac} (see also eqn.~\eqref{metpp}),
\bea
  2 ds^2 &=&
     d\mu^2+\tfrac{1}{4}\sin^2\mu(\sigma_1^2+\sigma_2^2)
        +\tfrac{1}{4}\sin^2\mu\cos^2\mu \,\sigma_3^2 \\
     && +\ \tfrac{1}{2}\left\{(\Sigma_1-\cos\mu\sigma_1)^2
        +(\Sigma_2-\cos\mu\sigma_2)^2
        +(\Sigma_3-{\textstyle \frac{1}{2}}(1+\cos^2\mu)\sigma_3)^2
     \right\} \nonumber \,,
\label{metpp1}
\eea
where $\Sigma_i$ are right-invariant one-forms on $SO(3)$, and
$\sigma_i$ are right-invariant one-forms on $SU(2)$. Let us introduce
coordinates so that
\bea \Sigma_1&=&\sin\phi
d\theta-\cos\phi\sin\theta d\psi\nn \Sigma_2&=&\cos\phi
d\theta+\sin\phi\sin\theta d\psi\nn \Sigma_3&=&d\phi+\cos\theta
d\psi \,,
\eea
with $0<\theta<\pi$, $0<\phi<2\pi$ and $0<\psi<2\pi$ so that we
parametrise $SO(3)$ rather than $SU(2)$. Note that we have scaled
the seven-dimensional metric so that $\Ricci=6 g$.

By defining $\psi'=-\psi/2$ and completing the square, the metric can be written as in (\ref{canform}) above
with the 6-metric on the flag manifold given by
\bea
ds^2_6&=&
\tfrac{1}{4}\left[
   d\theta-\cos\mu(\sin\phi \sigma_1+\cos\phi \sigma_2)\right]^2 \nn
&& +\tfrac{1}{4}\sin^2\theta\left[
   d\phi -\cos\mu\cot\theta(\cos\phi\sigma_1-\sin\phi\sigma_2)
   -\tfrac{1}{2}(1+\cos^2\mu)\sigma_3\right]^2\nn
&& +\tfrac{1}{2}\left[d\mu^2
     +\tfrac{1}{4}\sin^2\mu(\sigma_1^2+\sigma_2^2)
     +\tfrac{1}{4}\sin^2\mu\cos^2\mu\sigma_3^2\right] \,,
\eea
which displays it explicitly as an $S^2$ bundle over $\bCP^2$.
In addition we have the one-form $\sigma$ in (\ref{canform}) given
by
\be
\sigma=
-\tfrac{1}{2}\cos\mu\sin\theta(\cos\phi \sigma_1- \sin\phi\sigma_2)
+\tfrac{1}{4}\cos\theta(1+\cos^2\mu)\sigma_3
-\tfrac{1}{2}\cos\theta d\phi \,,
\ee
which indeed satisfies $d\sigma=2J_6$ where
\bea
   J_6 &=& \tfrac{1}{4}\sin\theta\left[
      d\theta-\cos\mu(\sin\phi\sigma_1+\cos\phi\sigma_2)\right] \nn
   && \quad \wedge\left[d\phi
      - \cos\mu\cot\theta(\cos\phi\sigma_1-\sin\phi\sigma_2)
      -\tfrac{1}{2}(1+\cos^2\mu)\sigma_3 \right] \nn
   && + \cos\phi\sin\theta K^1-\sin\phi\sin\theta K^2-\cos\theta K^3
\eea
and $K^i$ are the following basis for self-dual two-forms on
$\bCP^2$:
\bea
K^1&=&\tfrac{1}{4}\sin\mu d\mu\wedge \sigma_1+\tfrac{1}{8}\cos\mu\sin^2\mu\sigma_2\wedge\sigma_3\nn
K^2&=&\tfrac{1}{4}\sin\mu d\mu\wedge \sigma_2+\tfrac{1}{8}\cos\mu\sin^2\mu\sigma_3\wedge\sigma_1\nn
K^3&=&\tfrac{1}{4}\cos\mu \sin\mu d\mu\wedge \sigma_3+\tfrac{1}{8}\sin^2\mu\sigma_1\wedge\sigma_2 \,.
\eea

We now consider the 5-cycles $\Sigma_5$ on the seven-manifold defined by either
$\theta=0$ or $\theta=\pi$. These are $U(1)$ bundles over $\bCP^2$.
They are supersymmetric cycles provided that the corresponding 6-cycles
on the Calabi-Yau four-fold cone with base $N(1,1)_I$ are in fact
divisors. In our normalisations, the metric on the Calabi-Yau cone is
given by
\be
ds^2_8=dr^2+r^2[(d\psi'+\sigma)^2+ds^2_6] \,,
\ee
with the corresponding K\"ahler-form given by
\be
J_8=rdr\wedge (d\psi'+\sigma)+r^2 J_6 \,.
\ee
It can now easily be checked that the relevant six-cycles are indeed
calibrated by $\frac{1}{6}(J_8)^3$. M5-branes wrapped on these cycles
are expected to correspond to baryons in the dual SCFT. The conformal
dimension of these baryon operators will be given by the geometric
formula~\cite{gk,Fabbri:1999hw}
\be
\Delta=\frac{\pi N}{6}\frac{{\rm Vol}(\Sigma_5)}{{\rm Vol}(X_7)} \,.
\ee
In calculating the volume of the five-cycle with, say, $\theta=0$,
one first needs to shift the coordinate $\psi\to\psi-\phi$ to
ensure that the coordinates are well defined at $\theta=0$. Having
done this we find $\Delta=N/2$ which exactly agrees with that
expected from the description of the field theory dual to
$N(1,1)_I$ presented in~\cite{Billo:2000zr}. In particular the
operators $\det(U^i_\alpha)$ have conformal dimension $N/2$.

It would be interesting to determine the stability groups of the
supersymmetric cycles and then quantise the collective coordinates of
the wrapped five-brane, following the calculations
in~\cite{Witten:1998xy,Fabbri:1999hw}. This would give predictions for
the representations under the global symmetry group $SU(3)\times
SO(3)$ that the baryons carry and this could be compared with the
representations in $\det(U^i_\alpha)$ that are chiral primary.

\section{Parametrisations of $N(1,1)$}
\label{coord}

We have used two different sets of coordinates on $N(1,1)$. One,
following~\cite{Page:1984ac}, is the standard set where the manifold is
explicitly an $\SO(3)$ fibration over $\bCP^2$. The $N(1,1)_I$ and
$N(1,1)_{II}$ metrics in these coordinates are given in
eqn.~\eqref{metpp}. The other set, discussed in section~\ref{N11},
was adapted to the construction of $N(1,1)_I$ as a hyper-K\"ahler
quotient. In this appendix we show how these two parametrisations
are related. The easiest way to do this is to recall that
topologically
$N(1,1)\simeq\SU(3)/U(1)$. The two coordinate systems are then related
by two different parametrisations of $\SU(3)$.

Specifically $N(1,1)$ is defined to be the coset $SU(3)/U(1)$ where
the $U(1)$ is embedded in the $\l_8=\frac{1}{\sqrt{3}}{\diag}(1,1,-2)$
direction. Let us write a general $\SU(3)$ element as
\be
\label{uvt}
g = \begin{pmatrix}
      u^1 & -\bar{v}^1 & t^1 \cr
      u^2 & -\bar{v}^2 & t^2 \cr
      u^3 & -\bar{v}^3 & t^3
   \end{pmatrix} \in SU(3) \,.
\ee
For $(u^i,v_i)$, the $SU(3)$ condition implies that
$|u^i|^2=|v^i|^2=1$ and $u^iv_i= 0$. Identifying $(u^i,v_i)$ with the
corresponding variables in section~\ref{N11}, we see that these
conditions exactly match the hyper-K\"ahler quotient conditions
together with fixing the radius on the cone $C(N(1,1)_I)$. Once a
value of $(u^i,v_i)$ satisfying these constraints is chosen, the other
$SU(3)$ constraints involving $t^i$ can be solved completely by setting
$\bar{t}_i=-\epsilon_{ijk} u^j\bar{v}^k$. So, in the discussion below, we
regard $t^i$ as a function of $(u^i,v^i)$. In the hyper-K\"ahler
quotient we also modded out by the $U(1)$ action $(u^i, v^i)\sim
(e^{i\theta}u^i,e^{-i\theta}v^i)$. For consistency this must
correspond to the action of $\l_8=\frac{1}{\sqrt{3}}{\diag}(1,1,-2)$
on $g\in\SU(3)$. It is easy to see that this indeed that case.

The parametrization~\cite{Page:1984ac} of $N(1,1)$ as an $\SO(3)$
fibration over $\bCP^2$ is related to the well-known identification
$\bCP^2\simeq(SU(3)/\bZ_2)/U(2)$. To form this coset we write the
general $\SU(3)$ element (following the discussion in appendix~A
of~\cite{Cvetic:2001zx}) as
\begin{align}
   g
   &=
   \begin{pmatrix} \bar{a}_1 & \bar{a}_2 & 0 \cr -a^2 & a^1 & 0 \cr 0
      & 0 & 1 \end{pmatrix}
   \begin{pmatrix}\cos\m & 0 & \sin\m \cr 0 & 1 & 0 \cr -\sin\m & 0 &
      \cos\m \end{pmatrix}
   \begin{pmatrix} b^1 & -\bar{b}_2 & 0 \cr b^2 & \bar{b}_1 & 0 \cr 0 & 0 & 1
      \end{pmatrix}
   \begin{pmatrix} e^{i\theta} & 0 & 0 \cr 0 & e^{i\theta} & 0 \cr
      0 & 0 & e^{-2i\theta} \end{pmatrix}.
\label{mettransf}
\end{align}
where $a^\a$ and $b^\a$ form unit quaternions. The last factor is the
$U(1)$ subgroup used to form $N(1,1)\simeq\SU(3)/U(1)$. Together with
the second last factor it forms an $U(2)$ subgroup. Note that the
unit quaternion made of $b^\a$ actually parametrises
$S^3/\bZ_2\simeq\SO(3)$ since $(b^1,b^2)\sim(-b^1,-b^2)$ are
identified by the $U(1)$ element with $\theta=\pi$. The other unit
quaternion $a^\a$ and the real number $\m$ are coordinates on the
$\bCP^2$ base. Thus quotienting by $U(1)$ we see explicitly that
$N(1,1)$ is a $\SO(3)$ fibration over $\bCP^2$.

Comparing~\eqref{mettransf} to the parametrization~\eqref{uvt} we get
an explicit transformation between the coordinates on $N(1,1)$ used in
section~\ref{N11} and those of ref.~\cite{Page:1984ac}. In particular,
it is easy to show, writing $\sigma_i$ and $\Sigma_i$ for the
right-invariant one-forms constructed from $a^\a$ and $b^\a$,
respectively, that the two forms of the $N(1,1)_I$
metric~\eqref{metpp} and~\eqref{metuv2} are equivalent.


Let us end by noting that the $N(1,1)_I$ metric is {\em not} the same
as the usual coset metric obtained by simply gauging away the $U(1)$
factor from the standard $ds^2=-{\Tr}(g^{-1}dg)^2$ of $SU(3)$. The
latter in the notation of section~\ref{N11} is given by
\bea
ds^2 &=& d\bar{w}_i dw^i - |\bar{w}_i dw^i|^2
+  d\bar{z}^i dz_i-|\bar{z}^i dz_i|^2
+ d\bar{s}_i ds^i-|\bar{s}_i ds^i|^2
\nn
&&
+{\textstyle \frac{1}{2}}(d\phi^3 + i(\bar{w}_i dw^i + \bar{z}^i dz_i))^2
\eea
where $w^i$ and $z_i$ are coordinates on $\bCP^2$ satisfying
$w^iz_i=0$ while $\bar{s}_i=-\epsilon_{ijk}w^j\bar{z}^k$. This differs
from~\eqref{metuv2} by an extra $\bCP^2$-metric factor for $s^i$.



\begin{thebibliography}{99}


\bibitem{Maldacena:1997re}
  J.~M.~Maldacena,
  ``The large N limit of superconformal field theories and supergravity,''
  Adv.\ Theor.\ Math.\ Phys.\  {\bf 2} (1998) 231
  [Int.\ J.\ Theor.\ Phys.\  {\bf 38} (1999) 1113]
  [arXiv:hep-th/9711200].

\bibitem{Aharony:2002hx}
  O.~Aharony, B.~Kol and S.~Yankielowicz,
  ``On exactly marginal deformations of N = 4 SYM and type IIB  supergravity on
  $AdS(5)\times  S^5$,''
  JHEP {\bf 0206} (2002) 039
  [arXiv:hep-th/0205090].



\bibitem{Lunin:2005jy}
  O.~Lunin and J.~Maldacena,
  ``Deforming field theories with U(1) x U(1) global symmetry and their gravity
  duals,''
  arXiv:hep-th/0502086.


\bibitem{Gauntlett:2004zh}
  J.~P.~Gauntlett, D.~Martelli, J.~Sparks and D.~Waldram,
  ``Supersymmetric AdS(5) solutions of M-theory,''
  Class.\ Quant.\ Grav.\  {\bf 21} (2004) 4335
  [arXiv:hep-th/0402153].

\bibitem{Gauntlett:2004yd}
  J.~P.~Gauntlett, D.~Martelli, J.~Sparks and D.~Waldram,
  ``Sasaki-Einstein metrics on S(2) x S(3),''
  arXiv:hep-th/0403002.



\bibitem{Cvetic:2005ft}
  M.~Cvetic, H.~Lu, D.~N.~Page and C.~N.~Pope,
  ``New Einstein-Sasaki spaces in five and higher dimensions,''
  arXiv:hep-th/0504225.


\bibitem{Gursoy:2005cn}
  U.~Gursoy and C.~Nunez,
  ``Dipole Deformations of N=1 SYM and Supergravity backgrounds with U(1) X
  U(1) global symmetry,''
  arXiv:hep-th/0505100.


\bibitem{Klebanov:1998hh}
  I.~R.~Klebanov and E.~Witten,
  ``Superconformal field theory on threebranes at a Calabi-Yau  singularity,''
  Nucl.\ Phys.\ B {\bf 536} (1998) 199
  [arXiv:hep-th/9807080].


\bibitem{Acharya:1998db}
  B.~S.~Acharya, J.~M.~Figueroa-O'Farrill, C.~M.~Hull and B.~Spence,
  ``Branes at conical singularities and holography,''
  Adv.\ Theor.\ Math.\ Phys.\  {\bf 2}, 1249 (1999)
  [arXiv:hep-th/9808014].


\bibitem{Morrison:1998cs}
  D.~R.~Morrison and M.~R.~Plesser,
  ``Non-spherical horizons. I,''
  Adv.\ Theor.\ Math.\ Phys.\  {\bf 3}, 1 (1999)
  [arXiv:hep-th/9810201].




\bibitem{Gauntlett:2004hh}
  J.~P.~Gauntlett, D.~Martelli, J.~F.~Sparks and D.~Waldram,
  ``A new infinite class of Sasaki-Einstein manifolds,''
  arXiv:hep-th/0403038.







\bibitem{Fabbri:1999hw}
  D.~Fabbri, P.~Fre', L.~Gualtieri, C.~Reina, A.~Tomasiello, A.~Zaffaroni and A.~Zampa,
  ``3D superconformal theories from Sasakian seven-manifolds: New  nontrivial
  evidences for AdS(4)/CFT(3),''
  Nucl.\ Phys.\ B {\bf 577} (2000) 547
  [arXiv:hep-th/9907219].

\bibitem{Billo:2000zr}
  M.~Billo, D.~Fabbri, P.~Fre, P.~Merlatti and A.~Zaffaroni,
  ``Rings of short N = 3 superfields in three dimensions and M-theory on
  AdS(4) x N(0,1,0),''
  Class.\ Quant.\ Grav.\  {\bf 18} (2001) 1269
  [arXiv:hep-th/0005219].

\bibitem{Gukov:1999ya}
  S.~Gukov, C.~Vafa and E.~Witten,
  ``CFT's from Calabi-Yau four-folds,''
  Nucl.\ Phys.\ B {\bf 584}, 69 (2000)
  [Erratum-ibid.\ B {\bf 608}, 477 (2001)]
  [arXiv:hep-th/9906070].


\bibitem{Castellani:1983yg}
  L.~Castellani, L.~J.~Romans and N.~P.~Warner,
  ``A Classification Of Compactifying Solutions For D = 11 Supergravity,''
  Nucl.\ Phys.\ B {\bf 241} (1984) 429.




\bibitem{Duff:1986hr}
  M.~J.~Duff, B.~E.~W.~Nilsson and C.~N.~Pope,
  ``Kaluza-Klein Supergravity,''
  Phys.\ Rept.\  {\bf 130} (1986) 1.

\bibitem{Castellani:1998nz}
  L.~Castellani, A.~Ceresole, R.~D'Auria, S.~Ferrara, P.~Fre and M.~Trigiante,
  ``G/H M-branes and AdS(p+2) geometries,''
  Nucl.\ Phys.\ B {\bf 527} (1998) 142
  [arXiv:hep-th/9803039].



\bibitem{Witten:1981me}
  E.~Witten,
  ``Search For A Realistic Kaluza-Klein Theory,''
  Nucl.\ Phys.\ B {\bf 186} (1981) 412.



\bibitem{Castellani:1983mf}
  L.~Castellani, R.~D'Auria and P.~Fre,
  ``SU(3) X SU(2) X U(1) From D = 11 Supergravity,''
  Nucl.\ Phys.\ B {\bf 239} (1984) 610.









\bibitem{D'Auria:1983vy}
  R.~D'Auria, P.~Fre and P.~van Nieuwenhuizen,
  ``N=2 Matter Coupled Supergravity From Compactification On A Coset G/H
  Possessing An Additional Killing Vector,''
  Phys.\ Lett.\ B {\bf 136}, 347 (1984).



\bibitem{Cvetic:2001zx}
  M.~Cvetic, G.~W.~Gibbons, H.~Lu and C.~N.~Pope,
  Phys.\ Rev.\ D {\bf 65} (2002) 106004
  [arXiv:hep-th/0108245].


\bibitem{Page:1984ac}
  D.~N.~Page and C.~N.~Pope,
  ``New Squashed Solutions Of D = 11 Supergravity,''
  Phys.\ Lett.\ B {\bf 147} (1984) 55.

\bibitem{Castellani:1983tc}
  L.~Castellani and L.~J.~Romans,
  ``N=3 And N=1 Supersymmetry In A New Class Of Solutions For D = 11
  Supergravity,''
  Nucl.\ Phys.\ B {\bf 238} (1984) 683.







\bibitem{Gauntlett:2004hs}
  J.~P.~Gauntlett, D.~Martelli, J.~Sparks and D.~Waldram,
  ``Supersymmetric AdS backgrounds in string and M-theory,''
  arXiv:hep-th/0411194.


\bibitem{Chen:2004nq}
  W.~Chen, H.~Lu, C.~N.~Pope and J.~F.~Vazquez-Poritz,
  ``A note on Einstein-Sasaki metrics in $D\ge 7$,''
  arXiv:hep-th/0411218.


\bibitem{Ceresole:1999zg}
  A.~Ceresole, G.~Dall'Agata, R.~D'Auria and S.~Ferrara,
  ``M-theory on the Stiefel manifold and 3d conformal field theories,''
  JHEP {\bf 0003}, 011 (2000)
  [arXiv:hep-th/9912107].


\bibitem{Martelli:2004wu}
  D.~Martelli and J.~Sparks,
  ``Toric geometry, Sasaki-Einstein manifolds and a new infinite class of
  AdS/CFT duals,''
  arXiv:hep-th/0411238.


\bibitem{Gauntlett:2002fz}
  J.~P.~Gauntlett and S.~Pakis,
  ``The geometry of D = 11 Killing spinors.''
  JHEP {\bf 0304} (2003) 039
  [arXiv:hep-th/0212008].

\bibitem{Duff:1983nu}
  M.~J.~Duff, B.~E.~W.~Nilsson and C.~N.~Pope,
  ``Spontaneous Supersymmetry Breaking By The Squashed Seven Sphere,''
  Phys.\ Rev.\ Lett.\  {\bf 50}, 2043 (1983).





\bibitem{8dsugra}
  E.~Cremmer, B.~Julia, H.~Lu and C.~N.~Pope,
  ``Dualisation of dualities. I,''
  Nucl.\ Phys.\ B {\bf 523}, 73 (1998)
  [arXiv:hep-th/9710119].


\bibitem{Figueroa-O'Farrill:1999va}
  J.~M.~Figueroa-O'Farrill,
  ``On the supersymmetries of anti de Sitter vacua,''
  Class.\ Quant.\ Grav.\  {\bf 16} (1999) 2043
  [arXiv:hep-th/9902066].





\bibitem{Awada:1982pk}
  M.~A.~Awada, M.~J.~Duff and C.~N.~Pope,
  ``N = 8 Supergravity Breaks Down To N = 1,''
  Phys.\ Rev.\ Lett.\  {\bf 50} (1983) 294.



\bibitem{hart}
G.~W.~Gibbons, S.~A.~Hartnoll and C.~N.~Pope,
``Bohm and Einstein-Sasaki metrics, black holes and cosmological event
horizons,''
Phys.\ Rev.\ D {\bf 67} (2003) 084024
[arXiv:hep-th/0208031].





\bibitem{Duff:1984sv}
  M.~J.~Duff, B.~E.~W.~Nilsson and C.~N.~Pope,
  ``The Criterion For Vacuum Stability In Kaluza-Klein Supergravity,''
  Phys.\ Lett.\ B {\bf 139} (1984) 154.

\bibitem{Page:1984ad}
  D.~N.~Page and C.~N.~Pope,
  ``Stability Analysis Of Compactifications Of D = 11 Supergravity With SU(3) X
  SU(2) X U(1) Symmetry,''
  Phys.\ Lett.\ B {\bf 145}, 337 (1984).

\bibitem{Page:1984ae}
  D.~N.~Page and C.~N.~Pope,
  ``Which Compactifications Of D = 11 Supergravity Are Stable?,''
  Phys.\ Lett.\ B {\bf 144}, 346 (1984).


\bibitem{Englert:1982vs}
  F.~Englert,
  ``Spontaneous Compactification Of Eleven-Dimensional Supergravity,''
  Phys.\ Lett.\ B {\bf 119} (1982) 339.

\bibitem{Duff:1982ev}
  M.~J.~Duff,
  ``Supergravity, The Seven Sphere, And Spontaneous Symmetry Breaking,''
  Nucl.\ Phys.\ B {\bf 219} (1983) 389.



\bibitem{Englert:1983qe}
  F.~Englert, M.~Rooman and P.~Spindel,
  ``Supersymmetry Breaking By Torsion And The Ricci Flat Squashed Seven
  Spheres,''
  Phys.\ Lett.\ B {\bf 127}, 47 (1983).

\bibitem{Biran:1986zg}
  B.~Biran and P.~Spindel,
  ``New Compactifications Of N=1, D = 11 Supergravity,''
  Nucl.\ Phys.\ B {\bf 271}, 603 (1986).


\bibitem{Page:1984fu}
  D.~N.~Page and C.~N.~Pope,
  ``Instabilities In Englert Type Supergravity Solutions,''
  Phys.\ Lett.\ B {\bf 145} (1984) 333.






\bibitem{Ito:1984uj}
  K.~Ito,
  ``Instability Of Englert Solution On Squashed Seven Sphere In
  Eleven-Dimensional Supergravity,''
  Phys.\ Lett.\ B {\bf 147} (1984) 52.

\bibitem{Yasuda:1984vt}
  O.~Yasuda,
  ``Stability Of Englert Type Solutions On N(Pqr) In D = 11 Supergravity,''
  Phys.\ Rev.\ D {\bf 31} (1985) 1899.


\bibitem{Pope:1984bd}
  C.~N.~Pope and N.~P.~Warner,
  ``An SU(4) Invariant Compactification Of D = 11 Supergravity On A Stretched
  Seven Sphere,''
  Phys.\ Lett.\ B {\bf 150}, 352 (1985).

\bibitem{Pope:1984jj}
  C.~N.~Pope and N.~P.~Warner,
  ``Two New Classes Of Compactifications Of D = 11 Supergravity,''
  Class.\ Quant.\ Grav.\  {\bf 2}, L1 (1985).


\bibitem{AV}
C. Ahn and JF. Vazquez-Poritz,
``Marginal Deformations with $U(1)^3$ Global Symmetry'',
hep-th/0505168.

\bibitem{gk}
  S.~S.~Gubser and I.~R.~Klebanov,
  ``Baryons and domain walls in an N = 1 superconformal gauge theory,''
  Phys.\ Rev.\ D {\bf 58} (1998) 125025
  [arXiv:hep-th/9808075].


\bibitem{Witten:1998xy}
E.~Witten,
``Baryons and branes in anti de Sitter space,''
JHEP {\bf 9807} (1998) 006
[arXiv:hep-th/9805112].


\end{thebibliography}
\end{document}